\documentclass[12pt]{article}

\usepackage{amsmath,amsfonts,epsfig,color,latexsym}

\newcommand{\be}{\begin{equation}}
\newcommand{\ee}{\end{equation}}
\newcommand{\bea}{\begin{eqnarray}}
\newcommand{\eea}{\end{eqnarray}}

\begin{document}

\begin{flushright}
hep-th/0305069\\
BROWN-HET-1356
\end{flushright}
\vskip.5in

\begin{center}

{\LARGE\bf On Dp-Dp+4 systems, QCD dual and phenomenology }
\vskip 1in
\centerline{\Large Horatiu Nastase}
\vskip .5in

\end{center}
\centerline{\large Brown University}
\centerline{\large Providence, RI, 02912, USA}

\vskip 1in

\begin{abstract}

{\large D4-D8 and D3-D7 systems are studied and a possible holographic 
dual of large N QCD (SU(N) gauge fields and fundamental quarks) is 
sought. 
A candidate system is found, for which however
 no explicit solution is available. Susy is broken by having a 
$D7-\bar{D7}$ condensing to a D5. The mechanism for supersymmetry breaking is
then used to try to construct a Standard Model embedding. One can 
either obtain too few low energy fields or too many. The construction
requires TeV scale string theory.  
}

\end{abstract}

\newpage

\section{Introduction}

One of the reasons why people are interested in gravity-field theory
dualities is the hope that we can describe QCD via gravity. 
The original paper \cite{maldacena} treated the duality between ${\cal N}$=4 
SU(N) SYM at large N 
 and $AdS_5 \times S_5$ string theory, followed shortly thereafter 
by the paper of Witten \cite{wittenul} describing how to get a description of 
the pure glue theory (SU(N) Yang-Mills) via gravity. Other
developments afterwards involve breaking supersymmetry e.g. \cite{kasi} and/or 
conformal invariance e.g. \cite{ps,ks,malnu,malna}, and also introducing 
bifundamental fields via D7 branes or D5 branes e.g. \cite{kr,grapo,kk},
 but the question of having a theory without conformal invariance and 
supersymmetry and with dynamical massless quarks has eluded attempts 
to solve it. For a review of the developments until 1999 see \cite{agmoo}.
 In this paper I will address this question, and find
that while the decoupled D3-D7 theory describes a supersymmetric version of 
QCD, there is a modification of this system which describes the 
(large N) QCD, but unfortunately it is not possible to write down
explicitly (not even implicitly). 

The observation which allows us to do this is that one can write 
down the solution for a $D8-\bar{D8}$ system, and that if this system 
condenses to a D6, one can write that solution down as well. 
I use the embedding of massive 10d IIA theory in M theory defined 
in \cite{hull}
 and extended in \cite{lnr} to give a UV definition of the massive IIA 
solutions. It is then found that one has to T dualize and go to 
a D3-D7 system, smeared over an overall transverse coordinate. 
The holographic dual to QCD is then a system of $D3-(D7-\bar{D7}(D5))-
D7'$ branes, which however doesn't have a known supergravity solution.
The $D3-D7-\bar{D7}$ decoupled solution can be written explicitly (up to 
a one dimensional integral). The solution for the $D3-D7-\bar{D7}-D5$ can 
be found up to an integro-differential equation for a function $H_{4, p;q}(x)
$, but for the $D3-D7-\bar{D7}-D5-D7'$ even the ansatz cannot be written.

An obvious question then is can one lift this D brane construction for 
the holographic dual of QCD to a Standard Model embedding? I study 
this question in the context of D-brane-world GUT models and find that 
one needs to have TeV-scale string theory. In the context of an SU(5) 
susy GUT we can obtain massles states corresponding to the 5 fermions,
Higsses and gauge fields, but no $\bar{10}$ fermions (which contain 
the fields in the $(3,2)$ of $SU(3)\times SU(2)$). By adding
orientifolds, one is able to obtain the required fields, but much more
on top of that, and the corresponding masses seem to be wrong anyway. 
In any case, that system is very hard to analyze. 

The paper is organized as follows. In section 2 I study D4-D8
solutions, in 2.1. previous solutions and in 2.2 solutions which will 
be used thereafter. In section 3 I try to define the D4-D8 system and
its M theory embedding using the construction in \cite{hull,lnr}. Section 4 is 
devoted to motivating the supergravity- field theory correspondence 
for the D4-D8 system and identifying the gauged supergravity
describing it. In section 5.1 I describe the proposed set-up for the 
holographic dual of QCD, analyze the brane-antibrane condensation 
process and then in section 5.2 study the susy breaking using 
the $D8-\bar{D8}$ system. In section 6 I try to embed the 
QCD description into a  Standard Model description using D-brane 
worldvolumes. After
 a study of supersymmetric lagrangians for D3-D7-O(7) systems 
in 6.1, I try to build a model. In 6.2 systems without 
orientifolds are studied and section 6.3 introduces orientifolds. I finish in 
section 7 with discussion and conclusions. The appendix reviews 
GUTs for our purposes.

\section{D4-D8 systems and QCD}
From the perspective of a braneworld scenario or holographic 
duality, if one wants to realize a QCD-like system one has to introduce
fundamental quarks, and the most obvious way is to look at Dp-Dp+4 
systems. Since it is not clear how a D5-D9 system would be useful 
for either holography or braneworld phenomenology, we are left with 
D3-D7 systems and D4-D8. Let's start out with D4-D8 systems for 
their simplicity and then notice that one needs to go back to D3-D7, but 
keep the advantages of the D4-D8. 

The D4 brane theory dimensionally reduced to 4d is ${\cal N}$=4 SYM, 
which has fermions in the adjoint representationof $SU(N_c)$.
That theory is conformal, however in 5d the D4 theory has a
dimensionful coupling constant and is getting strongly coupled in the 
UV, therefore the theory is not well defined. We will get back later 
to the question of defining the theory. 
By introducing $N_f$ D8s we have quarks and scalars in the bifundamental
representation $(N_c, \bar{N_f})+(\bar{N}_c, N_f)$, forming an ${\cal
  N}$=2 hypermultiplet. In the holographic context, since 
\be
g^2_{Dp}=g_s l_s^{p-3}
\ee
in the decoupling limit $l_s\rightarrow 0$, keeping $g^2_{Dp}$ fixed 
means $g^2_{Dp+4}\rightarrow 0$, so we are left with $N_f$ fundamental
hypermultiplets. 

The D8 has the string metric and dilaton ($d \sigma_{8,1}^2$ is the
$8+1$-dimensional Minkowski metric)
\bea
ds^2&=& H^{-1/2} (d\sigma_{8,1}^2) + H^{1/2} dx^2 \nonumber\\
e^{\phi}&=& H^{-5/4} \nonumber\\
H&=& c+|\tilde{M}| |x|=c+\frac{m}{l_s}|x|
\eea
where $c$ is an arbitrary constant of integration or (by the usual 
rescaling for p-branes)
\bea
ds^2&=& \bar{H}^{-1/2}(d\bar{\sigma}_{8,1}^2)+\bar{H}^{1/2} d\bar{x}^2
=[\frac{3}{2}(1+\frac{g_sm}{l_s} |z|)]^{-1/3}(d\vec{\sigma}^2_{8,1}+dz^2) 
\nonumber\\
e^{\phi}&=& e^{\phi_0}\bar{H}^{-5/4}=g_s \bar{H}^{-5/4}
\nonumber\\
\bar{H}&=&H/c= 1+ g_s |\tilde{M}| |\bar{x}| =1+ \frac{g_s m}{l_s}|\bar{x}|
\label{deig}
\eea
where $g_s$ is defined as the coupling constant at the position of the 
D8 brane. Here
$\tilde{M}(x)=\pm H'$, so the mass is piecewise constant, and jumps at
the positions of the D8 branes. The $\pm$ in the mass corresponds to
D8 branes vs. anti-D8 branes. One can obviously redefine x such that 
the harmonic function appears just as a conformal factor for flat space.

\subsection{D4-D8 in the  literature}

The solution for a D4-D8 system can be written as 
\bea
ds^2&=& H_8^{-1/2} [H_4^{-1/2}d\vec{x}^2_{4,1} +H_4^{1/2}
d\vec{y}^2_4]+H_8^{1/2} H_4^{1/2} dz^2\nonumber\\
e^{\phi-\phi_0}&=& H_8^{-5/4}H_4^{-1/4}\nonumber\\
F_{(4)}&=&Q_4 vol(\Omega_{4,transv}) \nonumber\\
H_8(z)&=&1+g_s|\tilde{M}| |z|\nonumber\\
M&=&\pm H_8'
\label{deight}
\eea
and here also $\pm$ corresponds to the D8 vs. anti-D8.
In the literature, people have considered a ``D4 inside D8'' 
solution \cite{clp3} and a ``partially localized D4-D8 system
\cite{youm}. The ``D4 inside D8'' is just a D4 delocalized over the
transverse coordinates to the D8 and is given by (in Einstein frame) 
\bea
ds^2_E&=&W^{2/25}[H^{-3/8}_4(-dt^2 +d\vec{x}_4^2)+H^{5/8}_4
(dr^2 +r^2d\Omega_3^2)]+H^{5/8}_4dz^2\nonumber\\
 \hat{A}_{(3)}&=&\frac{\bar{Q}}{4m}W^{32/25}\Omega_3\nonumber\\
e^{\hat\phi}&=&W^{-4/5}H^{-1/4}\nonumber\\
H_4&=&1+\frac{\bar{Q}}{r^2}\;\;\;\;W=1+k|z|
\eea
Here $\bar{Q}$ is the $Q_4$ density on the unit of z direction. 
It was obtained by lifting the D4 in 9d solution via KK reduction 
on the D8 domain wall (that paper introduced the notion of dimensional 
reduction on a domain wall; by contrast one can always dimensioanlly
reduce on a coordinate paralel to the domain wall), with ansatz
\bea
d\hat{s}_{10, E}^2&=&e^{-\frac{5}{16}\sqrt{\frac{2}{7}}\phi}W^{\frac{2}{25}
}ds_{9,E}^2+e^{\frac{35}{16}\sqrt{\frac{2}{7}}\phi}dz^2\nonumber\\
\hat{A}_{(1)}&=&0,\;\;\;\; \hat{A}_{(2)}=\frac{1}{2m}W^{\frac{16}{25}}
F_{(2)},\;\;\;\; \hat{A}_{(3)}=\frac{1}{4m} W^{\frac{32}{25}}F_{(3)}
\nonumber\\
e^{\hat{\phi}}&=&W^{-\frac{4}{5}}e^{-\frac{7}{8}\sqrt{\frac{2}{7}}\phi}
\label{eq:ansatz}
\eea
The partially localized solution of Youm \cite{youm}reads
\bea
ds^2&=&\Omega^2(z)(H_4^{-1/2}(-dt^2+d\vec{x}^2_4)+H_4^{1/2}(d\vec{y}^2 
+dz^2))\nonumber\\
\Omega(z)&=&(\frac{3}{2}\frac{g_sm}{l_s}z)^{-1/6}\nonumber\\
e^{\Phi}&=&g_s(\frac{3}{2}\frac{g_sm}{l_s}z)^{-5/6}\nonumber\\
H_4&=&\frac{Q_4}{l_s^{10/3}(\vec{y}^2+z^2)^{5/3}}
\eea
and corresponds to the case $H_8=\frac{g_s m}{l_s}|\bar{x}|
=(\frac{3}{2}\frac{g_sm}{l_s}|z|)^{3/2}$ (that is, for 
$g_s\rightarrow \infty$, see (\ref{deig})).
Through a change of coordinates $z=r sin\alpha, y=r cos\alpha$ 
in was shown in \cite{bo}  that one finds a metric
\be
ds^2=(\frac{3}{2} Cm sin\alpha)^{-1/3}(Q_4^{-1/2}r^{4/3}dx_{||}^2
+Q_4^{1/2}\frac{dr^2}{r^2}+Q_4^{1/2}d\Omega_4^2)
\label{bo}
\ee
which is  a form where now $\alpha$ (compact coordinate, in
$S_1/Z_2$, since $0<\alpha <\pi/2$) can be
interpreted as being the transverse coordinate to the D8
due to the lucky coincidence that there is no r dependence in the 
transverse metric. Note that just for the D8 solution it would not 
be true, but since we have a D4 harmonic function with the correct 
r dependence, the total r dependence cancels. Because of that 
cancellation, (\ref{bo}) can 
be interpreted as the metric in the presence of an O(8) (and at 
infinite coupling; presumably there is a finite coupling O(8) one
could add to the D4-D8 system, and it should limit to this). 
One needs to add O(8) planes at the fixed planes to cancel the 
$D8$ brane charge (16 $D8$ branes to cancel -16 units of charge from two 
O8's). The authors of \cite{bo} found that the corresponding dual 
$D4$-$D8$ theory is a fixed point  with global $SU(2)\times E_{N_f+1}$
global symmetry, derived from a $Sp(Q_4)$ gauge theory  theory at 
infinite bare coupling. So, although the D4 theory is not conformal, 
by adding the D8 and O(8), the theory flows to a nontrivial conformal 
fixed point.

\subsection{D4-D8 solutions}

However, finding a localized solution is not so difficult after all.
The point is that there are so called partially localized intersections,
where brane 1 with harmonic function $H_1$ 
lives on $t, \vec{w}, \vec{x} $, and brane 2 with harmonic function
$H_2$ lives on 
$t, \vec{w}, \vec{y}$, with overall transverse space $\vec{z}$. They
are written in terms of harmonic functions $H_1$ and $H_2$ in the
usual way, except that now $H_1$ and $H_2$ 
satisfy the equations (e.g. \cite{youm}, \cite{lupo})
\bea
\partial_z^2 H_1(z, y) + H_2(z) \partial_y^2 H_1(z, y)=0, \partial_z^2
H_2=0 \;\; {\rm or}\nonumber\\
\partial_z^2 H_2(z, x) + H_1(z) \partial_x^2 H_2(z, x)=0, \partial_z^2 
H_1=0
\eea
In other words, we delocalize one brane (say brane 2) over the
worldvolume coordinates of the other brane (1), 
and then $H_1$ is harmonic  (obeys the laplace equation) 
in the background of brane 2. In the case of a Dp-Dp+4 system, this 
condition is automatically satisfied, and then $H_4$ obeys the equation
\be
\partial_z^2 H_4 (z, \vec{y}) +H(z) \partial_y^2 H_4(z,
\vec{y})=Q\delta (z) \delta^4(\vec{y})
\label{eqhar}
\ee
where $Q=a g_s N l_s^3$, with $a$ some numerical constant, and then 
\be
H_4(y,z)=1+\int \frac{d^4 p}{(2\pi)^4}e^{i\vec{p}\vec{y}}H_p(z)
=1+\frac{1}{4\pi^2 y}\int _0^{\infty}dp p^2 J_1(py)H_p(z)
\ee
If we put $H(z)=c+m |z|$, then the resulting equation 
\be
H_p''(z)-(c+m|z|)p^2 H_p(z)=Q\delta(z)
\label{diff}
\ee
is solved by 
\be
H_p(z)=c_p \bar{z}^{1/2} K_{1/3}(\frac{2}{3} \bar{x}^{3/2}), \;\; 
\bar{x}=(\frac{p}{m})^{2/3} (c+m |z|)
\ee
The constant $c_p$ is fixed by matching with the normalization of
the $\delta $ function source, and one gets 
\be
c_p = \frac{Q \sqrt{c}}{2p^{1/3} m^{2/3} [K_{1/3} ( \frac{2}{3}
  \frac{p}{m} c^{3/2})- \frac{p}{m}c^{3/2}K_{4/3} (\frac{2}{3}
\frac{p}{m} c^{3/2})]}
\ee
and so 
\be
H_4(y,z)=1+\frac{Q\sqrt{c}}{4\pi^2 m^{2/3}}\frac{\beta^{1/3}}{y}
\int dp p^2 \frac{J_1(py) K_{1/3} (\frac{2}{3} \beta p)}
{K_{1/3} ( \frac{2}{3}
  \frac{p}{m} c^{3/2})- \frac{p}{m}c^{3/2}K_{4/3} (\frac{2}{3}
\frac{p}{m} c^{3/2})}
\ee
with 
\be
\beta =  \frac{(c+m|x|)^{2/3}}{m}
\ee
Let us now define the decoupling limit. As I mentioned, 
we want to keep the D4 SYM coupling fixed, 
$g^2_{D4}=g_sl_s$, so that 
\be
H_8=1+\frac{g_s N_f}{l_s}|z|= 1+g^2_{D4} N_f |Z|={\rm fixed}
\ee
(I have rescaled as usual $z=l_s^2 Z$  and so $m=M/l_s^2$ 
and the number of D8's is $N_f$. 
Then by rescaling also 
$y=l_s^2 U$ and the integration variable $p=P/l_s^2$, we get 
in the limit $l_s\rightarrow 0$
\be
H_{D4}(U,Z)\simeq \frac{1}{l_s^4}\frac{a g^2_{D4} N}{4\pi\tilde{\alpha}^{2/3}}
\frac{\bar{\beta}^{1/3}}{U}\int_0^{\infty} dP P^2 \frac{
J_1(PU) K_{1/3}(\frac{2}{3} \bar{\beta} P)}{K_{1/3}
(\frac{2}{3}\frac{P}{\tilde{\alpha}}) -\frac{P}{\tilde{\alpha}}
K_{4/3}(\frac{2}{3}\frac{P}{\tilde{\alpha}})}
=\frac{h_{D4}(U,Z)}{l_s^4}
\label{harmonic}
\ee
and the decoupled D4-D8 system is 
\be
ds^2=\alpha ' H_8^{-1/2}(Z)[ h_{D4}^{-1/2}(U,Z)dx_{||}^2 +
h_{D4}^{1/2}(U,Z) (H_8(Z)dZ^2 + dU^2 +U^2 d\Omega_3^2)]
\label{decoupled}
\ee
A similar analysis for the case of the D2-D6 system was done in 
\cite{cha} and for a D1-NS5 system it was done in \cite{lnr}. 
Notice that the near core D8 can be always trusted, independent 
of the number of D8 branes (the curvature in string units
 is always small, curvature scalar $R\sim M^2H^{-5/2}$, so one needs 
actually $g_s N_8\ll 1$, which can be satisfied for $N_f \sim 1$).
The number of D4's however, has to be very large, as usual. 

Let us now look at various ways of breaking supersymmetry. 
The most used is the method of Witten \cite{witten2}, of putting 
the system at finite temperature. This corresponds to 
compactifying on a supersymmetry breaking circle. The fermions
aquire masses of the order of the compactification scale at 
tree level, and the scalars at quantum level, by fermion loops.

For the holographic dual, Witten's solution had the AdS black hole
as a starting point. Then 
scale the mass M to infinity, together with r to infinity and t to 0
in a particular way. The resulting solution has only one parameter (the 
radius of AdS). 

But equivalently \cite{hr} one can just take the near
horizon limit of the nonextremal solution. Although this solution has
apparently two parameters (the AdS radius and the nonextremality 
parameter M, or temperature T), calculations in this background will 
not depend on T alone. Indeed there exists a rescaling (with no 
parameters going to infinity!) which takes the D3 nonextremal 
near-horizon solution to Witten's metric, namely $U=\rho (TR), t=t'/(TR),
\vec{y}=\vec{x}/T$ (R=AdS radius). In particular, for the Wilson loop
\cite{maldac} 
calculation of $q\bar{q}$ potential in \cite{bisy}, this means that
E(L,R,T)=E'(LT,R) (TR) or equivalently EL=f(LT,R). The bottom line is 
that when one computes either $q\bar{q}$ potential or glueball masses, 
 one can use either Witten's type of 
construction, or a near horizon nonextremal solution, in which case by 
scaling of the coordinates one gets the desired nonsusy theory. Both 
ways were used in glueball calculations \cite{coot,mjmn,cort,ct,bmt}, but when 
one starts with a nonconformal theory before the compactification, 
there is no analog of the AdS black hole solution (since there is no AdS 
background), so one has to use the scaling in the nonextremal solution.
A similar case, of glueballs in the N=1 nonconformal cascade theory of
Klebanov and Strassler \cite{ks}, was treated in \cite{ch}.

So let us try to put the D4-D8 system at finite T by making it 
nonextremal. It is easy to do so for the ``D4 inside D8'' solution.
Just lift the solution for D4 in 9d at finite temperature on the D8 
with the ansatz (\ref{eq:ansatz}) and get 
\be
ds^2=H_8^{-1/2}[H_4^{-1/2}(-dt^2f(y)+dx_{||}^2)+H_4^{1/2}
(\frac{dy^2}{f(y)}+y^2 d\Omega_3^2+H_8dz^2)]
\ee
where 
\be
f(y)=1-\frac{\mu}{y^2}
\ee
But this is again the delocalization over z of the full nonextremal 
D4-D8 solution,
which however now is hard to find. 

If one could find the finite temperature localized D4-D8 solution,
it would still not be so useful, since the corresponding 
field theory will be the same as for pure D4 branes at finite 
temperature: pure 4d Yang-Mills theory. Maybe though by comparing 
the two descriptions one would be able to find out the spurious 
effects of the construction (by seeing if there are quantities which 
do change). 

We want however to keep the fundamental fermions after susy breaking. 
When compactifying the D4-D8 field theory, we would like therefore to 
put antiperiodic boundary conditions for the (4,4) fermions, so that
they become massive, and periodic boundary conditions for the 
(4,8) fermions, so that they remain massless.  One would have to 
check whether such boundary conditions are consistent with the
interactions, and whether unitarity is preserved in such a system. 

But let's see whether one can find a holographic dual to such a
 system. 
 By the general argument in \cite{witten3}, if one dimensionally 
reduces on a euclidian black hole spacetime, all the fermions will be 
antiperiodic around the KK coordinate, so they will get a mass.
The argument is that there is only one spin structure available 
to the spinors around the KK coordinate. In general, the possible 
phases around it are dictated by  the invariances of the lagrangian.
At large distances from the black hole, the space is topologically 
flat (times the KK circle), so that all phases are allowed. Near the 
horizon however, the spacetime is flat space times the transverse sphere
and admits a unique spin structure, which becomes the antiperiodic 
one at infinity. The same argument can be extended to nonextremal 
branes, for instance nonextremal $D4$ branes, as in \cite{witten2}. 
The spacetime is flat at infinity and has a transverse sphere near 
the horizon. Only the antiperiodic spin structure is valid over the
whole spacetime.

So in the case of the euclidian black hole or nonextremal D4, fermions 
defined over the whole space become massive. They couple to fermionic 
operators on the D4. Therefore the D4 fermions are antiperiodic and 
become massive. 

But we also see a way out. If there are fermions which are defined
only over a part of the holographically dual spacetime, they can 
be mapped to fermionic operators remaining in the spectrum, whereas 
the ones defined over the whole spacetime are mapped to operators 
dissappearing from the spectrum. 

In the context of compactification, there doesn't seem to be a
solution of this type, but there is a domain wall type
solution (``alternative to compactification'') 
which has the necessary properties.

The nice thing about D8 branes is that one 
can also write down a $D8-\bar{D8}$ solution (unlike for other
branes), in the particular case where in between the two branes we
have flat space. This can be achieved by writing 
\bea
H_8&=& 1-2g_s|\tilde{M}|z, \;\;\; z<0,\;\;\; M=-g_s\tilde{M} \nonumber\\
&& 1, \;\;\; 0< z< z_0 \nonumber\\
&& 1+2g_s|\tilde{M}|(z-z_0), \;\;\; z_0<z \;\;\; M=-g_s\tilde{M}
\label{ddbar}
\eea
If one would have $M=+g_s\tilde{M}$ for $z>z_0$ it would be a D8-D8,
but now it is a $D8-\bar{D8}$, and if the D8 can be trusted
-curvature small in string units means 
\be
g_s N_8 \ll 1
\ee
and  $g_s$ small implies no quantum corrections-
the $D8-\bar{D8}$ can be trusted as well. 
All we did was change the sign of the mass on one
side, which changes the sign in the Killing spinor equation, therefore 
the Killing spinor on one side is not valid on the other. So there is 
no globally defined fermion in this background. 

One can still write down a localized D4 inside the D8, even in the
presence of the $\bar{D8}$. Piecewise, the $H_4$ equation is still
(\ref{diff}), with c and m derived from (\ref{ddbar}), and then we 
just have to match the solutions over the branes. 

In the case $z_0\rightarrow 0$, we have a $D8-\bar{D8}$ on top of 
each other. From the string theory point of view, 
that gives a gravitational solution, but can also
(depending on the K theory class of the system) give a lower
dimensional brane, e.g. a D6. From the gravity point of view, 
the holographic dual is the 
same as for the D4-D8, just that now $M=-g_s \tilde{M}$ on both sides 
of the brane. 

I will postpone the discussion of a specific set-up for later, but 
let us note that whereas there is no globally defined fermion, there
are fermions defined on the D8 (and at x=0 we still have
supersymmetry)-or rather on the z=0 slice of the holographic dual 
(\ref{decoupled}), corresponding to the D8. 

So in the case of the $D4-D8-\bar{D8}$ solution, fermions defined over 
the whole spacetime couple to field theory operators which will dissappear 
from the spectrum. These will be operators with no D8 charges. On the other 
hand, fermions defined only on the D8 will couple to fermionic operators 
in the field theory which remain in the spectrum. These are operators 
charged under the D8 global symmetry. 

Note that the fact that there are fermions which are defined only 
on a subset of the holographic dual is not a new concept. The ${\cal N}$=2
superconformal theory of D3-D7-O(7) described in \cite{afm} has 
bulk modes coupling to operators with no $N_f$ charges and 
vector modes defined on an $AdS_5 \times S_3$ orientifold fixed 
plane inside $AdS_5\times S_5$ coupling, to operators with $N_f$ (vector)
charges. It is in fact very similar to this system: without the O(7) and after 
a T duality it becomes the D4-D8, so the coupling of the operators with 
$N_f$ charges to vector modes defined on D8 (or rather the z=0 slice of 
(\ref{decoupled})) is an established  fact. The only new observation is that 
therefore the uncharged fermionic operators dissappear from the spectrum 
(get high anomalous dimensions), whereas the charged fermionic operators 
don't.

\section{Defining the D4-D8 system}

As I mentioned, the D4 field theory (and the D8 field theory
 actually, but that 
is now ``frozen'') is not well defined in the UV, so one must allow 
for a UV completion. In the UV, the effective D4 coupling is large, and the 
theory must be described by M theory, therefore the UV completion 
of the (nonrenormalizable) D4 theory is given by the M5 brane 
field theory. But what about the D4-D8 case? 

Let us start with seeing how to embed the D8 in M theory. There 
have been attempts to embed the D8 directly in M theory, as an 
``M9'' domain wall. One of these
is solution in \cite{bs}, where the 11th direction is an isometry 
direction for the metric. But the D8 is a solution to Romans' massive
supergravity \cite{romans2}, 
and a fully covariant M9 would be also a solution to an 
11d supergravity with a mass parameter (cosmological constant). 

It is unkown how to lift the massive IIA theory and its D8 background 
solution directly into M theory. The point is that M theory doesn't
seem to admit a cosmological constant, and if one wanted to lift massive 10d
supergravity directly, one would get a cosmological constant in 11d. 
The only possible way around this is if the 10d mass arises via a 
Scherk-Schwarz generalized reduction on a circle. But for that, one 
would need a global symmetry in 11d, and the action doesn't have such
a symmetry. The equations of motion have however a scaling symmetry,
which was exploited in \cite{lalupo} to reduce to a massive 10d sugra.
However, that is a different massive supergravity in 10d (one that
admits, in particular, the de Sitter space as a background), and
moreover, it amounts in 11d to a compactification on the euclidian
radial direction. That massive supergravity can also be obtained as a 
usual reduction of a modified M theory, as in \cite{hlw}.

Instead, the most conservative embedding in M theory was 
done by Hull \cite{hull}, who 
was able to embed the massive supergravity and the D8 background
in M theory by introducing two extra T dualities, one of which was a
``massive T duality'' as defined in \cite{brgpt}. M theory on a $T^2$
of zero area is type IIB and IIB can be compactified on $S^1$ via 
Scherk-Schwarz. After a ``massive T duality,'' we get massive IIA.

The endpoint is that  massive IIA
supergravity is equivalent to M theory on the space B(A,R), in the
limit $A\rightarrow 0, R\rightarrow 0$, with  metric 
\be
ds_B^2= R_3^2 (dx_3)^2 +\frac{A}{Im (\tau)}|dx_1 +\tau(x_3)x_2|^2
=R_3^2 (dx_3)^2 + R_2^2(dx_2)^2 + R_1(dx_1+ m x_3 dx_2)^2
\label{space}
\ee
and all the radii going to zero, and the x's have periodicity 1, $x_i
\sim x_i+1$. In the limit, we should keep the massive IIA quantities 
fixed, so 
\be
g_s=\frac{l_s}{Im(\tau_0)R_3}=\frac{R_1 l_s}{R_2 R_3}= {\rm fixed},
\;\; l_s =\frac{l_P^{3/2}}{R_1^{1/2}}={\rm fixed}\;\;  ({\rm and \;\;
m \;\; fixed})
\ee

So how does the Hull duality help us in defining the D8 and the 
D4-D8 field theories?
The correct description of the D8 brane field theory is not clear, but 
the D8 goes over to a gravitational background in the (Hull) 
dual M theory, so the field theory on that soliton in M theory will 
be the correct description. 

In the D4-D8 case, the D8 field theory decouples, but one is left with (4,8)
fields, which can't be lifted to usual M theory, so one needs to look 
for the Hull dual.  The field theory description is given by the 
lift of the D4-D8 system to M theory. One is T dualizing twice to 
get to M theory, so it matters where are located the dualized
coordinates. There are 3 choices: T dualize along two transverse 
coordinates, two paralel coordinates, or along one each. The latter 
situation is the most useful, since then after two T dualities, one is 
still describing a D4 brane, albeit with two small radii. 
The D8 background has now become a 6-brane smeared over two transverse
directions. Then the
lift to M theory gives as usual a M5 brane, in the background 
dual to the D8 (i.e. a 7+1-dimensional worldvolume, described in
(\ref{mtheory})). The decoupling of the 
D8 field theory corresponds to decoupling of the degrees of freedom 
localized at the M theory 7+1d background, but one still has degrees of 
freedom coming from M2's stretched between the M5 and the ``M7''
(corresponding to (4,8) strings). 
Unfortunately, it is unclear how to describe these degrees
of freedom, but at 
least it is possible in principle. 
In the next subsection I will analyze in more detail the embedding 
in M theory via Hull duality.

\subsection{Embedding in M theory }

Let us now derive the embedding of supergravity solutions of massive IIA 
into 11d supergravity solutions (if the solutions are BPS, the
embeddings are still valid, even if the space is singular, and one
can't use quantum perturbation theory). 

 Dimensionally reducing 
M theory to massless IIA  on $r_1$ one has
\bea
ds_{11}^2&=& e^{-2\phi/3}(ds_8^2 +r_2^2 A^2 dz_2^2+ r_3^2 B^2 dz_3^2) + 
 e^{4\phi /3}r_1^2 (dz_1+ A_2 dz_2 +A_{\mu}dz^{\mu})^2
\nonumber\\
C_{(3)}&=&A_{(3)}+dz^1\wedge B_{(2)} 
\eea

Now one has to perform a T duality on $z_2$ to get to IIB and then a 
massive T duality on $z_3$ to get to massive IIA.

The full set of T duality rules giving the (hatted) massless IIA 
fields in terms of the IIB ones  are 
\bea
\hat{g}_{00}&=&\frac{1}{g_{00}}\;\;\;
\hat{g}_{0i}= \frac{B_{0i}}{g_{00}}\;\;\;
\hat{g}_{ij}= g_{ij}-\frac{g_{0i}g_{0j}-B_{0i}B_{0j}}{g_{00}}
  \nonumber\\
\hat{B}_{0i}& =& \frac{g_{0i}}{g_{00}}\;\;\;
\hat{B}_{ij} = B_{ij}+ \frac{g_{0i}B_{0j}-B_{0i}g_{0j}}{g_{00}}
\nonumber\\
\hat{\phi}&=& \phi-\frac{1}{2} log (g_{00})\nonumber\\
\hat{A}_{ijk}&=&\frac{8}{3}D^+_{0ijk}+B_{0[i}B_{jk]}^{(2)}-B^{(2)}_{0[i}B_{jk]}
+B_{0[i}B^{(2)}_{|0|j}g_{k]0}/g_{00}-B^{(2)}_{0[i}B_{|0|j}g_{k]0}/g_{00}
\nonumber\\
\hat{A}_{0ij}&=& \frac{2}{3}[B^{(2)}_{ij}+2\frac{B^{(2)}_{0[i}g_{j]0}
}{g_{00}}]\nonumber\\
\hat{A}_i&=&-B^{(2)}_{0i}+a B_{0i}\;\;\;
\hat{A}_0= a
\eea

The inverse T duality rules, this time with the added complication of
them being massive, are (this time the hatted quantities are IIB and 
unhatted massive IIA)
\bea
\hat{g}_{00}&=&\frac{1}{g_{00}}\;\;\;
\hat{g}_{0i}= \frac{B_{0i}}{g_{00}}\;\;\;
\hat{g}_{ij}= g_{ij}-\frac{g_{0i}g_{0j}-B_{0i}B_{0j}}{g_{00}}
  \nonumber\\
\hat{B}_{0i}& =& \frac{g_{0i}}{g_{00}}\;\;\;
\hat{B}_{ij} = B_{ij}+ \frac{g_{0i}B_{0j}-B_{0i}g_{0j}}{g_{00}}
\nonumber\\
\hat{\phi}&=& \phi-\frac{1}{2} log (g_{00})\nonumber\\
\hat{D}^+_{0ijk}&=&\frac{3}{8}
[ A_{ijk}-A_{[i}B_{jk]}+\frac{g_{0[i}B_{jk]}
A_0}{g_{00}}-\frac{3}{2} \frac{g_{0[i}A_{jk]0}}{g_{00}}
\nonumber\\
\hat{a}&=& A_{0}+mx^0\nonumber\\
B^{(2)}_{ij}&=&\frac{3}{2}A_{ij0}-2A_{[i}B_{j]0}+2\frac{g_{0[i}B_{j]0}A_0}{
g_{00}}+mx^0(B_{ij}+2g_{0[i}B_{j]0}g_{00})\nonumber\\
B_{0i}^{(2)} &=& -A_i+\frac{A_0g_{0i}}{g_{00}}
\eea

Applying the above T duality rules going from IIA to IIB on $z_2$ and 
then to massive IIA on $z_3$ one gets for the 11d metric (keeping 
only the fields relevant for our discussion)
\bea
ds_{11}^2 &=& e^{-2\phi/3}(ds_8^2 +r_3^2 B^2 dz_3^2)
\nonumber\\&&+e^{4\phi/3}r_1^2
[(dz_1+adz_2+B_{\mu 2}^{(2)}dz^{\mu})^2+e^{-2\hat{\phi}+2\hat{\phi}_0}
dz_2^2]\nonumber\\
&=&e^{-2\phi/3}(ds_8^2 +r_3^2 B^2 dz_3^2)\nonumber\\
&&+e^{4\phi/3}r_1^2
[(dz_1+(m+ A_3)dz^2 +A_{\mu 23}dz^{\mu})^2 +e^{-2\phi}A^2
\frac{r_2^2}{r_1^2} dz_2^2]
\eea
and the corresponding massive IIA metric is then 
\be
ds_{10 mA}^2 =ds_8^2 +A^{-2}\frac{dz_2^2}{r_2^2} + B^{_2}
\frac{dz_3^2}{r_3^2}
\ee
while the dilaton is ($\hat{\hat{\phi}}$ is the massive IIA dilaton, 
$\hat{\phi}$ is the IIB dilaton and $\phi$ the massless IIA dilaton)
\be
e^{\hat{\hat{\phi}}}=\frac{e^{\hat{\phi}}}{Br_3}=\frac{e^{\phi}}{AB}
\frac{r_1}{r_2r_3}
\ee
When one applies this prescription to the D8 solution of massive type 
IIA 
\bea
ds^2_{10mA}&=& H^{-1/2} (d\sigma_{8,1}^2) + H^{1/2} dx^2 \nonumber\\
e^{\phi}&=& H^{\frac{-5}{4}} \nonumber\\
 H&=& c+|\tilde{M}| |x|=c+\frac{m}{l_s}|x|
\eea
one indeed finds the 11d gravitational metric
\be
ds^2_{11}= H^{1/2}(H^{-1/2} d \vec{\sigma}^2_{6,1} +H^{1/2} dx^2) +ds_B^2
=d \vec{\sigma}^2_{6,1} +H dx^2 +ds_B^2
\label{mtheory}
\ee
with
\be
ds_B^2= H (r_3^2dx_3^2 +r_2^2dx_2^2) +\frac{r_1^2}{H}(dx_1 + mx_3
dx_2)^2
\ee
Hull \cite{hull} has also found this solution as the correct 11d
gravitational background corresponding to the D8 background. We should 
note here that the massive IIA sugra does not admit flat space as a solution, 
the background with maximal supersymmetry is the D8.

Now when one lifts a solution of massive IIA to M theory, 
it matters where one chooses to make the two T dualities, i.e. where 
one puts $z_2$ and $z_3$. The best choice is of course to arrange $z_2$
and $z_3$ such as to get the same type of solution after 
the two T dualities. 

In the particular case of the D4-D8 solution, the best choice is to
have one direction paralell to the D4, one perpendicular. Then after 
two T dualities, one still has the D4 solution, and it will lift to 
an M5. 

Let us however first treat 
 the case where both T dualities are paralel to the 
D4. We will reach a D2 which lifts to an M2 in the gravitational 
background. Indeed, one gets 
\bea
ds^2_{10mA}&=& H_8^{-1/2}[H_4^{-1/2}(d\vec{\sigma}^2_{2,1}+\frac{dz_2^2}{
r_2^2}+\frac{dz_3^2}{r_3^2})+H_4^{1/2} d\vec{r}_4^2 ] +H_8^{1/2}
H_4^{1/2}dx^2 \nonumber\\
e^{\hat{\hat{\phi}}}&=& H_8^{-5/4}H_4^{-1/4}
\eea
which implies
\bea
ds_{11}^2 &=& H_4^{-2/3}d\vec{\sigma}^2_{2,1}+ H_4^{1/3}[d\vec{r}_4^2
\nonumber\\
&&+H_8(dx^2+r_3^2 dz_3^2 + r_2^2 dz_2^2)+H_8^{-1}r_1^2 (dz_1+mz_3
dz_2)^2]
\eea
Restricting the D4-D8 to the ``D4 inside D8'' solution
 corresponds as before just to dropping the x
dependence of $H_4$. 

When one T duality is paralel and one perpendicular, the same solution 
as above, but with $z_3$ a transverse coordinate, lifts to 
\bea
ds_{11}^2 &=& H_4^{1/3} (d\vec{\sigma}^2_{3,1}+H_8 r_3^2 dz_3^2 
+H_8^{-1}r_1^2(dz_1+mz_3dz_2)^2)\nonumber\\&&
+H_4^{2/3}(d\vec{r}_3^2
+H_8(dx^2 +r_2^2 dz_2^2))
\eea
which corresponds to an M5 in the gravitational background (\ref{mtheory}).
Note that in both cases there is also a nontrivial $F_{(4)}$ field.

In \cite{lnr}, a procedure was developped for getting a Matrix model
\cite{bfss,grt,wat} 
corresponding to the massive IIA supergravity, and was applied to 
the D8 background. I will apply it now to the D4-D8 system. 

Since massive 10d IIA string theory is equivalent to M theory 
on the singular background (\ref{mtheory}), one defines Matrix theory in that 
background and compactifies it. After T dualities in all the $r_i$, 
one gets a Matrix model of D3 branes. As an intermediate step necessary 
to decouple gravity from the D3 brane theory, following 
Sen \cite{sen} and Seiberg \cite{seiberg}, an $\bar{M}$ theory 
was introduced, such that 
\be
\frac{R_s}{\bar{l}_P^2}=\frac{R}{l_P^2}, \;\; \frac{\bar{R}_i}{
\bar{l}_P}=\frac{R_i}{l_P}
\ee
are held fixed in the $\bar{l}_P \rightarrow 0$ limit, and the $l_P 
\rightarrow 0$ limit is imposed afterwards. 

The metric (\ref{space}) for B(A,R) is invariant under the isometries
(we have put $R_i=1$ for simplicity)
\bea
T_1:&& x_1\rightarrow x_1+a_1, x_2\rightarrow x_2, x_3\rightarrow
x_3\nonumber\\
T_2:&&x_2\rightarrow x_2+a_2, x_1\rightarrow x_1, x_3\rightarrow
x_3\nonumber\\
T_3:&&x_3\rightarrow x_3+a_3, x_1\rightarrow x_1-m x_2a_3, 
x_2\rightarrow x_2
\eea
with Killing vectors $V_1=\partial _1$, $V_2=\partial_2$ and $V_3=
\partial_3-m x_2\partial_1$. One also notes that $[T_2,T_3]\neq 0$.
Since $T_2$ and 
$T_3$ don't commute, it matters in which order one makes the T
dualities. We choose to do $T_1$, then $T_2$, then $T_3$. 

After $T_1$ one has  : 
\bea
ds^2&=& (dx_3^2 +  dx_2^2 
+ dx_1^2)\nonumber\\
B_{12}&=& m x_3 \rightarrow H_{123}=m \nonumber\\
e^{\phi}&=& e^{\phi_0}
\label{aftert1}
\eea
After $T_2$ one has 
\bea
ds^2 &=& (dx_3^2 + dx_1^2) + (dx_2 - m
x_3 dx_1)^2 \nonumber\\
e^{\phi}&=& e^{\phi_0} 
\label{aftert2} 
\eea
$T_3$ is generated by the vector 
$V_3=\partial_3 +m x_1\partial_2$. Making this $=\partial'_3$
so that one can apply the T duality rules means the coordinate transformation 
\be
x_3'=x_3, x_2'=x_2 + m x_1x_3
\label{coordtr}
\ee
The metric in the new  coordinates is (after dropping primes on
coordinates) 
\bea
ds^2 &=& ( dx_3^2 + dx_1^2) + (dx_2  + m
x_1 dx_3  )^2 \nonumber\\
e^{\phi}&=& e^{\phi_0} 
\label{aftercoord} 
\eea
After the third $T$-duality, we have 
\bea
ds^2 &=& dx_1^2  +  \frac{(dx_2^2
  + dx_3^2  )}{ 1 + m^2 x_1^2 }    \nonumber\\
B_{23}dx^2 \wedge dx^3&=& - \frac{m x_1}{1+ m x_1^2 } dx_2 \wedge dx_3
 \nonumber\\
e^{\phi} &=& \frac{e^{\phi_0}}{ (1+ m x_1^2 )^{1/2}}
\label{aftrt3}
\eea

So let us apply this procedure for the D4-D8 solution. One
goes to an $\bar{M}$ theory to 
decouple string theory, compactifies on a lightcone coordinate, and 
then T dualizes on all 3 $r_i$'s. 

We will drop the bars from all quantities (in the end, nothing will 
depend on the $\bar{M}$ theory anyway). Let's start with the 
background corresponding to an M2. The IIA metric after
dimensional reduction on the lightlike coordinate will be (we choose 
that coordinate to be perpendicular to the M2, thus getting a D2
brane)
\bea
ds^2&=& H_4^{-1/2}d\vec{\sigma}^2_{2,1}+H_4^{1/2}[d\vec{r}_3^2
|H_8(dx^2 +r_3^2 dz_3^2 +r_2^2 dz_2^2)
\nonumber\\&& +H_8^{-1}r_1^2(dz_1+mz_3
dz_2)^2]\nonumber\\
e^{\phi}&=&g_sH_4^{1/2}
\eea
After the T duality on $T_1$ it will become a D3 brane ending on 
a NS5 brane in the $z_1$ direction. The metric is 
\bea
ds^2&=& H_4^{-1/2}d\vec{\sigma}^2_{2,1}+H_4^{-1/2}
H_8\frac{dz_1^2}{r_1^2}+H_4^{1/2}d\vec{r}_3^2+H_4^{1/2}H_8[dx^2 +r_3^2
dz_3^2 +r_2^2 dz_2^2] \nonumber\\
e^{\hat{\phi}}&=&\frac{e^{\phi_0}}{r_1}H_8^{1/2}\nonumber\\
B_{12}&=& mz_3 \rightarrow H_{123}=m
\eea
After a T duality on $T_2$ we get a D4 brane again
\bea
ds^2&=&H_4^{-1/2}d\vec{\sigma}^2_{2,1}+H_4^{-1/2}H_8\frac{dz_1^2}{r_1^2}
+H_4^{1/2}d\vec{r}_3^2\nonumber\\
&&+H_4^{1/2}H_8(dx^2+r_3^2dz_3^2)+H_4^{-1/2}H_8^{-1}
\frac{1}{r_2^2}(dz_2+mz_3 dz_1)^2\nonumber\\
e^{\phi}&=& \frac{e^{\phi_0}}{r_1}H_4^{-1/4}
\eea
and finally after  the coordinate transformation and 
T duality on $T_3$ one gets a D5 brane (coming from the 
original D4) in the ``7-brane'' background
(corresponding to the original D8).  
\bea
ds^2&=&
H_4^{-1/2}d\vec{\sigma}^2_{2,1}+H_4^{-1/2}H_8\frac{dz_1^2}{r_1^2}
+H_4^{1/2}d\vec{r}_3^2+H_4^{1/2}H_8dx^2\nonumber\\
&& +H_4^{-1/2}H_8^{-1}\frac{dz_2^2/r_2^2+dz_3^2/r_3^2}{1+H_4^{-1}
H_8^{-2}\frac{m^2z_1^2}{r_2^2r_3^2}}\nonumber\\
e^{\phi}&=& \frac{e^{\phi_0}}{r_1r_2r_3}H_8^{-1/2}H_4^{-1/2}
[1+H_4^{-1}H_8^{-2}\frac{m^2z_1^2}{r_2^2r_3^2}]^{-1/2}
\nonumber\\
B_{23}dz^2 \wedge dz^3&=& -\frac{m r_1}{r_2r_3H_8^2}\frac{z_1/r_1}{
1+ H_8^{-2}(m \frac{r_1}{r_2r_3})^2 z_1^2/r_1^2} dz_2/r_2 \wedge dz_3
/r_3
\eea
String theory in the D4-D8 background corresponds to D3 Matrix theory
in the above (D5- ``7-brane'') background.
Note that one has a transverse ``lightcone'' coordinate, which really 
means that we have boosted the M2 in a transverse direction, so 
one has to approach the $R_s\rightarrow 0$ limit with care. 

Let us now analyze what happens in M theory when one decouples the 
D4-D8 theory. 

The decoupling limit of the D4-D8 is $l_s\rightarrow 0$, 
with $g_s l_s=g^2_{D4}$= fixed. But we have been a bit cavalier about 
the $l_s$ dependence. In the massive IIA metric, the radii are 
$l_s^2/r_i$, i=2,3, and are supposed to go to infinity (or be very
large). Taking $r_i=l_s^2 R_i$, $R_i$ are still very
small (though finite). In order for this to be  a good decoupling
limit in M theory, it is clear that $r_1$ has to be treated as $r_2,
r_3$ namely $r_1=l_s^2 R_1$ also, and one indeed finds that. Then
\be
g^2_{D4}=g_sl_s=\frac{l_P^3}{r_2r_3}=\frac{R_1}{R_2R_3}
\rightarrow l_P^3 \sim r_2 r_3\sim l_s^4
\ee
and as usual $\vec{r}_4=l_s^2 \vec{U}_4, x=l_s^2 X$
\be
H_4\sim \frac{g_s l_s^3}{r^3}\sim \frac{1}{l_s^4 U^3}\sim
\frac{1}{l_P^3}
\ee
Therefore the decoupling limit of the string theory corresponds to a 
decoupling limit of the M theory, with metric (in the M2 case)
\bea
ds_{11}^2&\sim& l_P^2 [ h_4^{-2/3}d\vec{\sigma}_{2,1}^2+h_4^{1/3}[
d\vec{U}_4^2+H_8(dX^2 +R_2^2 dz_2^2 +R_3^2 dz_3^2 )
\nonumber\\&&+H_8^{-1} R_1^2
(dz_1+mz_3dz_2)^2]]
\label{decoupledm}
\eea
Since in the decoupling limit the D4-D8 field theory is equivalent 
to string theory in the background (\ref{decoupled}) with harmonic 
function (\ref{harmonic}), by Hull duality the 
corresponding M2- ``M7'' field theory (``M7'' is the gravitational 
background (\ref{mtheory})) is dual to M theory in the background
(\ref{decoupledm}). 

In the M5 case, one would get a M5- ``M7'' duality in a similar manner.

In this analysis, depending on the position of the $z_2,z_3$ and lightcone
coordinate R (paralel or transverse to the M theory brane
corresponding to D4), one has different endpoints. One other possible 
complication arises if one wants to compactify this system to get to 
a 4d field theory. Then one would need to choose that coordinate v as
well.

Nondecoupled D4-D8 theory we saw  can be lifted in 3 ways to M theory
(M2, M5 or KK monopole always in the gravitational background
(\ref{decoupled})), which in turn 
can be described by a decoupled theory of D3 branes (describing 
gravity), with some added
branes (describing the D4-D8 background). 
I have analyzed the case of M2 with perpendicular
lightcone coordinate. Another case is obtained
if one chooses $x_{11}$  (the lightcone coordinate) and $z_2$ in D4 
and $z_3$ perpendicular to it and so lift to an M5 brane. Going to 
string theory on $x_{11}$, one gets a D4 with $z_1, z_2$ paralel and 
$z_3$ perpendicular to it. After
all 3 T dualities, one gets a D3' with only $z_3$ paralel, whereas the 
D3s coming from the Matrix D0 branes have paralel $z_1,z_2,z_3$. The (4,8) 
strings are mapped to strings going from D3' to the 8d plane
transverse to x and $x_1$ (``7-brane'').

So I have mapped the D4-D8 system to $D3\perp D3'(1)$ in the presence
of a 7-brane. Gravity comes from the moduli space of the D3 and one
still has the full field theory on D3. Decoupling of the gravity
corresponds to decoupling of the D3 theory, and one is left with the 
D3' theory, with string modes ending on the 7-brane. One has therefore  
described the decoupled D4-D8 theory by a decoupled 
D3- ``7-brane'' theory, which seems consistent, since the T dual 
to the D4-D8 should be a D3-D7.  At
an intermendiate step, the 11d description of the decoupled theory 
was in terms of a M5- ``M7'' theory, so the D4 theory was correctly 
lifted to the M5 theory, and there is an implicit (even if not 
useful) definition of the UV completion of the D4-D8 field theory. 

Finally, just as a curiosity,
let us see how far can we go in writing the metric for 
the holographic dual of the D3 theory giving the Matrix model 
in the presence of the D5 and ``7-brane'' (the M2 picture 
for string theory in the D4-D8 background). The simplest place to 
start is after T duality on $T_1$, when the Matrix model will be in 
terms of D1 branes in the background of D3s ending on smeared NS5s.
D1 is in t and $z_1$ direction, D3 in $\vec{\sigma}_{2,1}$ and $z_1$, 
and NS5 in $\vec{\sigma}_{2,1}$ and $\vec{r}_3$. The coordinates 
$z_1,z_2,z_3$ are smeared over 
and x is overall transverse. The solution is written
in the obvious way, but it satisfies the criterion of partially 
localized multiple intersections (since with the smearing in $z_1\perp
NS5$ we have $D1\in D3$ inside smeared NS5). Then the harmonic
functions obey the equations
\bea
&&\partial_x^2 H_5(x)=0\nonumber\\
&&\partial_x^2 H_3(x, \vec{r}_3)+H_5(x)\partial_{\vec{r}_3}^2H_3(x, 
\vec{r}_3)=0\nonumber\\
&&\partial_x^2 H_1(x, \vec{r}_3, \vec{\sigma}_2)+H_5(x) 
\partial_{\vec{r}_3}^2H_1(x, \vec{r}_3, \vec{\sigma}_2)+
H_3(x,\vec{r_3})
\partial_{\vec{\sigma}}^2H_1(x, \vec{r}_3, \vec{\sigma}_2)=0
\label{three}
\eea
To obtain the holographic dual one would need to make the $T_2$ T
duality, coordinate transformation and $T_3$ T-duality, and then take 
a decoupling limit. But for that one would need an explicit solution of
(\ref{three}), and one can solve just the first two equations 
(similar to (\ref{eqhar})), and then the third (the equation 
for $H_1$) is too complicated to solve. 

By defining
\bea
H_3(\vec{r}_3,x)&=& 1+\int \frac{d^3p}{(2\pi )^3} e^{i\vec{p}\vec{r}_3}H_{p;3}
(x)= 1+\frac{1}{(2\pi )^2 r_3}\int _0^{\infty} dp p sin (p r_3)H_{p;3}
(x) \nonumber\\
H_1(\vec{r}_3, \vec{\sigma}_2, x) &=& 1+\int \frac{d^3 p}{(2\pi )^3}
\int \frac{d^2q}{(2\pi)^2} e^{i\vec{p}\vec{r}_3+i\vec{q}\vec{\sigma}}
H_{p,q ;1}(x)\nonumber\\
&=&
 1+\frac{1}{(2\pi )^2 r_3}\int _0^{\infty} dp p sin (p r_3)
\int \frac{d^2q}{(2\pi)^2} e^{i\vec{q}\vec{\sigma}}H_{p,q;1}(x)
\eea
and setting $H_5(x)=c+m|x|$ one gets the equations
\bea
&& H_{p;3}''(x)-(c+m|x|)p^2 H_{p;3}(x)=Q_3\delta (x)
\nonumber\\
&&H_{p,q;1}''(x)-(c+m|x|)p^2 H_{p,q;1}(x)\nonumber\\
&&-q^2(H_{p,q;1}(x)+\int
\frac{d^3 p'}{(2\pi)^3}H_{|\vec{p}-\vec{p}'|,q;1}(x)H_{p';3}(x))
=Q_1\delta (x)
\eea
The first one is the same as (\ref{diff}), so is solved in the same way, 
but the second one is too hard.

\section{DW-QFT for D4-D8 and motivating the field theory-supergravity 
correspondence}

In this section I will analyze the field theory- gravity
correspondence for the D4-D8 system. I would  like to argue that 
the correct supergravity description of the D4-D8 field theory is 
in terms of a 6d ${\cal N}$=2 supergravity.

First notice that the decoupled D4-D8 solution in (\ref{decoupled}) 
is not of the $AdS_n
\times S_m$ type. It is not even a fibration of $AdS_6$ over $S_4$ 
like the $D4$-$D8$ solution of Brandhuber and Oz in (\ref{bo}), hence one will 
not have a corresponding conformal field theory, but rather a 
nonconformal quantum field theory. The theory in \cite{bo} became 
conformal due to the presence of the orientifold planes, together with 
the near-horizon limit. It was conjectured in \cite{bo} that the 
nonlinear KK reduction on the $AdS_6$ fibered over $S_4$ ($AdS_6\times 
S_4$ with a warp factor)
\be
ds_{10}^2=(sin \alpha)^{-1/3}(ds^2_{AdS_6}+ const.(d\alpha^2+
(cos\alpha)^2d\Omega_3^2))
\ee
will give the N=4 F(4) gauged sugra in 6d of Romans \cite{romans}, with 
an susy $AdS_6$ ground state and an SU(2) gauge group. The conjecture 
was later proven in \cite{clp2}. The F(4) sugra is the only N=4 gauged 
sugra in 6d with and $AdS_6$ ground state. But there are other N=4 
gauged sugras in 6d with no $AdS_6$ ground state. In particular, there 
is an SU(2) gauged sugra which arises as an $S_1$ reduction of the 
(minimal) N=2 gauged sugra in 7d \cite{tn}, with an SU(2) gauge group
and a topological mass term.
I will try to argue that this is the theory on the supergravity 
side of the correspondence. It is a good guess since the D4-D8 
system has ${\cal N}$=2 susy and so has SU(2) R-symmetry, related 
by the correspondence to the sugra gauge group. 

Let us see what kind of supergravities are available, and start in 7d,
where one knows the dual to the M5 field theory.
In seven dimensions there is a maximal (N=4) gauged sugra, with 
gauge group SO(5). It is obtained as a KK reduction of 11d sugra on $S_4$
\cite{nvv} and as such it gives the  gravity dual to the M5 brane
(the flat brane corresponds to its $AdS_7$ vacuum solution).
Then there are the minimal (N=2) gauged supergravity with SU(2) 
gauge group of \cite{tn} and the coupled N=2 sugra+ vector multiplet,
with gauge group SO(4) (SU(2) in the sugra multiplet and another SU(2)
in the vector multiplet) of \cite{salsez}.
 They were obtained as KK reduction
of 10d N=1 sugra on $S_3$ in \cite{nv},\cite{cs} and 
\cite{clp5} and as such 
give the gravity dual of the NS5 brane theory. The pure SU(2) sugra 
can be written using a 2-form field or its 3-form field dual, but it is
in the latter formulation only that one can add a ``topological mass'' 
term $h\epsilon_{(7)}dA_{(3)}A_{(3)}$, which can be made supersymmetric. 
Since the SO(4) sugra+matter can only be written in the 2-form 
formulation, it doesn't admit a topological mass term deformation.
By truncation of the maximal gauged sugra, one can obtain an SU(2)
gauged sugra with a fixed topological mass term $h$ (related to the 
gauge coupling $g$). The maximal sugra does contain a sugra multiplet 
and a vector multiplet, but has also a topological mass term, hence 
it cannot be consistently truncated to the SO(4) sugra. 

Going down to six dimensions, one has several gauged sugra models 
too. First, there is the dimensional reduction of the N=4 model, which 
gives an N=8 gauged sugra with SO(5) gauged group, written in
\cite{cowdall}. It is natural therefore to asociate it with the 
theory on a $D4$ brane, as was done indeed in \cite{bst}. 
If one dimensionally reduces the pure SU(2) 7d gauged theory with 
topological mass h, one generates an N=4 sugra with gauge group SU(2) 
coupled to an 
U(1) vector multiplet as in \cite{gpn}. When h=0 one can consistently 
truncate the vector, resulting in pure N=4 gauged sugra. Since the pure 
N=4 sugra
theory is also a consistent truncation of the maximal N=8 theory in 6d,
it should also be related to the $D4$ brane theory, but with half the 
supersymmetry. But there is yet another gauged SU(2) N=4 sugra, found 
by Romans \cite{romans}, which is just a different mass deformation of 
the pure case, not involving any new fields, but with a new parameter m. 
The Romans theory admits a supersymmetric ground state with the full 
$AdS_6$ symmetry group, F(4), if the gauge
coupling is related 
to the mass m by g=3m. Having F(4) as symmetry group, it is 
not surprising that it was found in \cite{bo}
to  correspond to the conformal field theory on a $D4$+$D8$+O8 system. 

Let us now match supergravities with brane systems. The M5 corresponds
to the 7d SO(5) supergravity, obtained by $S_4$ reduction of 11 d
sugra, and by a further $S_1$ reduction one relates the 6d SO(5) sugra
to the D4 theory. The type IIA NS5 brane theory is matched to the 7d SO(4)
sugra, by reduction on $S_3$ of 10d type IIA sugra (with the type I
NS5 subset related to the 7d SU(2) sugra by reduction of the 10d type
I sugra. The 7d SU(2) vector multiplet couples to operators charged 
under the vector multiplet on the IIA NS5). The 6d SU(2) sugra  
corresponds to a D4 with half the supersymmetry, that is to a 
D4 of type I in 9d (M5 corresponds to 7d SO(5) sugra, and we
compactify on a transverse circle, thereby modifying the transverse 
sphere as  $S_4 \rightarrow S_3 \times S_1$
and a paralel circle, giving in the end a D4 in 9d). 

The deformation with mass parameter $m=g/3$ in 6d gives the F(4) sugra, 
corresponding to the D4-D8-O(8) system. It is not clear to what corresponds
the deformation with $m$ independent of $g$, but maybe it means 
moving away from the orientifolds (away from the conformal point). The 
mass parameter $m$ is related by $S_4$ reduction of the massive 10d
sugra \cite{clp2} to the 10d mass, which in turn is 
related to the number of $D8$ branes by  (as used in \cite{bo})
$m \propto (8-N_f)$, where the 8 comes from the O8 charge. So the 6d 
mass deformation with parameter $m$ corresponds to adding O8 and $N_f$
$D8$ branes.

The deformation with mass parameter $h$ in 7d correponds to a
deformation of the type I NS5 brane theory, but a deformation 
outside the massless type 
IIA NS5. The only possibility is that the h deformation corresponds
to the mass deformation of 10d sugra, that is, to the NS5 paralel to
D8 system. (The NS5-D8(5) is a particular case of the ``overlapping 
brane system'' NS5-Dq(q-3), $3\leq q \leq 8$ \cite{gauntlett,ett},
 which has the M theory 
solution $M5\perp M5(1)$ as a prototype). 

This would mean in particular that the massive type IIA sugra 
in 10d compactified on $S_3$ would reduce to $h$-deformed 7d SU(2) sugra, 
with $h$ related to the 10d mass. While this is not proven, it seems very 
likely given the precedent: the massive 10d IIA compactified on $S_4$
gives the 6d F(4) sugra, with $g=3m$ related to the 10d mass.  
Yet another argument is the fact that  
11d sugra on $S_4$ can be consistently truncated to the SU(2) sugra 
with a fixed $h$ (proportional to $g$) \cite{nv}. 
The NS5-D8 solution will be obtained in a manner completely analogous 
to the D4-D8 solution (\ref{decoupled}). One can easily do it as an exercise.
Compactifying 10d sugra to 7d on the decoupled NS5-D8 solution should
give the same h-deformed sugra. On the other hand, compactifying 
NS5-D8 on an $S_1$ paralel to NS5 one gets a D4-D7 solution of massive 
9d sugra (which could be oxidized back to a D4-D8), related to 
h-deformed 6d SU(2) sugra. 

Let's mention now that in the nonconformal cases discussed (like 
the D4 and the D4-D8 systems) one has to turn to the so-called
Domain Wall-Quantum
Field Theory correspondence (DW-QFT), rather than AdS-CFT, a
particular case of gravity-field theory correspondence where the
backgrounds for sugra are domain walls rather than Minkowski or
(anti) de Sitter. 

The nonconformal cases of D-branes were studied first in \cite{imsy},
and later in \cite{bst}, where the term ``Domain Wall-QFT 
correspondence'' was coined. The reason for the name is that the authors
of \cite{bst} realized that the near-horizon solution of nonconformal 
D-branes in Einstein frame gives known domain wall solutions of 
gauged supergravities. These domain wall solutions become in the dual p-brane
frame just $AdS_n \times S_m$ vacua, and the correspondence becomes 
simpler (in particular, the UV-IR relation becomes just $E\sim u$).
Further treatments of the DW-QFT correspondence can be found in 
\cite{bh,clp4,clp}.

In particular, the analysis revealed that the $D4$ brane theory is dual 
to type IIA sugra on $S_4$, which has as a massless mode a N=8 (maximal)
 SO(5)- gauged sugra which was later \cite{cowdall} obtained as an $S_1
$ reduction of the maximal (N=4) SO(5)-gauged sugra in 7d. 
It was also found that the compactification on tori transverse to the 
brane produces as effective sugras ones with noncompact gaugings,
for $T^k$ reduction of the M2 brane one has a CSO(8-k,k) gauged N=8,
D=4 theory, for a $T^k $ reduction of the M5 brane  one has a CSO(5-k,k)
gauged D=7 sugra and for a $T^k$ reduction of the D3 brane one has a 
CSO(6-k,k) gauged D=5 sugra (the C stands here for contraction, which 
eliminates unphysical gauge fields of negative metric from the spectrum).
In particular, the D=5 CSO(6-k,k) series was later found in \cite{acfg},
whereas the D=7 CSO(5-k,k) series is still not constructed.

Finally, let us note that the h=0 6d SU(2) model should not only be a 
truncation of the 6d SO(5) model.
It should also be obtained as an alternative 1/2 susy truncation
of a CSO(4,1) model in 6d. The CSO(4,1) model in 6d is the circle 
reduction of the corresponding model in 7d, and corresponds to the 
near-horizon theory of the $D4$ brane in N=2, D=9 theory (circle reduction
of the usual type IIA $D4$ brane in D=10). 

In conclusion, one expects a DW-QFT correspondence to relate the 
6d SU(2)-gauged sugra, coupled with a mass parameter $h$ to an U(1) 
vector multiplet, to a D4-D8 system. The decoupled solution 
(\ref{decoupled}) should dimensionally reduce on $S_3$ to a 
solution of $h$-deformed 7d sugra. 

Unfortunately, just the $S^3$ reduction of the massive IIA theory would 
be a hard task, one that merits a whole new paper.

\section{Towards a holographic dual for QCD}

In this section I would like to take what we have learned about 
D4-D8 systems and see if one can write down a holographic dual for QCD.
The D4-D8 system contains all the fields necessary for QCD, so one 
needs to generate mechanisms for getting rid of the extra fields.

\subsection{Set-up, condensation}

First, one needs to break supersymmetry. As 
we saw, one way was to compactify with susy breaking conditions for
the fermions, which corresponds on the gravity dual side to making the 
solution nonextremal, but this kills unfortunately all 
fermions. Another way was to introduce a $\bar{D8}$, which 
was very easy to do in supergravity. 
Just flip the sign of the mass on one side of 
the $D8$, and you get a $D8-\bar{D8}$ background. The 
string theory $Dp-\bar{Dp}$ 
system is however not necessarily purely gravitational, but depending on the 
charges, can contain also lower (Dp-2, Dp-4,...) branes. So after the 
$D8-\bar{D8}$ condensation, one could be left with a D6 (One would not
like to be left with another D4, since then that D4 field theory will 
become again relevant, i.e. will not decouple. 
The D6 field theory is still decoupled). The solution 
for D4 inside $D8-\bar{D8}$ does not yet holographically 
describe the QCD-like field theory. A priori it is an unstable point in 
the dynamics, but we will argue that it should take a very large time to 
decay. However, the $D8-\bar{D8}$ by itself doesn't have any overall gauge 
fields left, so we need a lower brane to take the role of D8 in 
the dual. After the condensation has ended a D6 will be formed, so it is 
natural to represent it in the holographic dual even if the $D8-\bar{D8}$ 
is still there. It is understood as a core around which the condensation 
will eventually take place.

In conclusion, one would like to have both the $D8-\bar{D8}$ and the D6 
in the gravity solution, the $D8-\bar{D8}$ since the condensation should 
take infinite time, and the D6 since we want bifundamental fields.

If a $Dp$ brane ($p\le 6$) collides with a $\bar{Dp}$ brane, we expect
to form as an intermediate stage a black uncharged p-brane 
(the extension of the Schwarzschild
solution to a p+1 dimensional worldvolume), which might then decay to 
a Dp-2 brane, but the gravitational description of this string process
seems hard to obtain. But as we saw, 
in the particular case of p=8 there is no collapse
of the gravitational 
solution when the Dp touches the $\bar{Dp}$.
Even if the $D8-\bar{D8}-D6-D4$ solution is still unstable on a very large 
timescale, it will be 
holographically related 
shortly to massless QCD, so it is what we want. The issue of black
holes and understanding the susy breaking process 
will be postponed for the next subsection.  

Let us try to write down this D4 parallel to 
D6 parallel to $D8-\bar{D8}$ solution. 
It can be obtained implicitly, since the 
harmonic functions should satisfy (again invoking the criterion of 
partially localized multiple interesections)
\bea
&&\partial_x^2 H_8(x)=0\nonumber\\
&&\partial_x^2 H_6(x, \vec{r}_2)+H_8(x)\partial_{\vec{r}_3}^2H_6(x, 
\vec{r}_2)=0\nonumber\\
&&\partial_x^2 H_4(x, \vec{r}_2, \vec{\sigma}_2)+H_8(x) 
\partial_{\vec{r}_2}^2H_4(x, \vec{r}_2, \vec{\sigma}_2)+
H_6(x,\vec{r}_2)
\partial_{\vec{\sigma}_2}^2H_4(x, \vec{r}_2, \vec{\sigma}_2)=0
\eea
Solving it would be identical to solving (\ref{three}), so the 
discussion is the same. Thus the solution is 
\bea
H_8(x)&=&c+m|x|\nonumber\\
H_6(x,\vec{r}_2)&=&
1+\frac{1}{(2\pi )^2 r_2}\int _0^{\infty} dp p sin (p r_2)H_{p;6}
(x) \nonumber\\
&=&1+\frac{Q\sqrt{c}}{4\pi^2 m^{2/3}}\frac{\beta^{1/3}}{r_2}
\int dp p \frac{sin(pr_2) K_{1/3} (\frac{2}{3} \beta p)}
{K_{1/3} ( \frac{2}{3}
  \frac{p}{m} c^{3/2})- \frac{p}{m}c^{3/2}K_{4/3} (\frac{2}{3}
\frac{p}{m} c^{3/2})}\nonumber\\
H_4(x, \vec{r}_2,\vec{\sigma}_2)&=&
 1+\frac{1}{(2\pi )^2 r_2}\int _0^{\infty} dp p sin (p r_2)
\int \frac{d^2q}{(2\pi)^2} e^{i\vec{q}\vec{\sigma}_2}H_{p,q;4}(x)
\label{nondec}
\eea
and where the equation for $H_{p,q;4}(x)$ is 
\bea
&&H_{p,q;4}''(x)-(c+m|x|)p^2 H_{p,q;4}(x)\nonumber\\
&&-q^2(H_{p,q;4}(x)+\int
\frac{d^3 p'}{(2\pi)^3}H_{|\vec{p}-\vec{p}'|,q;4}(x)H_{p';6}(x))
=Q_1\delta (x)
\eea
So what field theory does this solution (or rather its decoupling
limit) describe? Since one still has the D6 brane 
in place of the D8, one still has the bifundamental fields, but there 
are now differences. 

First of all, susy is broken at the string scale (by the 
$\bar{D8}$), so the 
(4,4) (and (8,8)) adjoint fermions get a string scale mass. 

One comment is in order here. This statement doesn't imply any 
knowledge of nonperturbative physics, it is just meant to parametrize 
our ignorance. At the string scale we don't have control anymore (in 
particular, the $\bar{D8}$ might evaporate, say), so it is natural to put 
the susy breaking scale there (at least at the string scale would be more 
appropriate). 

But note that there still is supersymmetry on the D8 (at x=0), so if 
fermions are defined only on the D8, they will remain massless. 
This is indeed the case for the (4,8) (bifundamental) fermions.
Why is this so? There are no globally defined fermions in the 
bulk, and that translates to string-scale mass for the bulk fermions.
By the AdS-CFT, via the coupling of the closed string bulk modes 
to the (4,4) bilinear operators, the (4,4) fermionic operators will 
get a very large anomalous dimension and decouple. But the fermions 
defined on the D8 couple to fermionic bilinear operators with (4,8) 
quantum numbers ($SO(2N_f)$ quantum numbers), and these fermions 
will remain massless, therefore the (4,8) fermionic operators remain 
in the theory. This still does not say anything yet about the 
dynamics of the theory, just that the starting point has fundamental 
fermions, and no adjoints. The dynamics would be derived from the 
final decoupled holographic dual.

But in order to have a holographic dual for QCD, 
one still has to say what happens to the scalars (adjoint and
bifundamental), how to get rid of the fermions in the conjugate 
representation (we need ${\cal N}$=1 fermions, not ${\cal N}$=2), 
and also how to get to 4d. 

One should note here that an ${\cal N}$=1 scalar superfield has a 
complex scalar, related to two transverse coordinates, and one complex
fermion. 

To get to 4d, it would seem that one needs to compactify one coordinate.
But one can easily check (by looking at the D4 holographic dual 
analyzed in \cite{imsy}) that if one compactifies (on a small radius)
the D4 in 10d to a D3 in 9d and still insist on the decoupling of
gravity, one is forced to go to the T dual description of 
the holographic dual, namely the near-horizon D3 brane 
in 10d  with one transverse coordinate compactified. 
Indeed, the compactified D4 holographic dual is 
\bea
ds^2 &=& l_s^2 [\frac{U^{3/2}}{\sqrt{g^2_{D4}N}}(dx_{3+1}^2 +
R^2 dx_5^2)+\frac{\sqrt{g^2_{D4}N}}{U^{3/2}}(dU^2+U^2 d\Omega_4^2)]
\nonumber\\
e^{\phi}&=&[\frac{U^{3/2}g^6_{D4}}{N}]^{1/4}\nonumber\\
g^2_{D3}&=& g^2_{D4}/R
\eea
To get the same from the D3 brane in 10d,
\be
ds^2=
l_s^2[\frac{\bar{U}^2}{\sqrt{g^2_{D3}N}}dx_{3+1}^2+\frac{d\bar{U}^2
+\bar{U}^2d\Omega_5^2}{\bar{U}^2}]
\ee
with $\bar{r}=l_s^2 \bar{U}, r=l_s^2U$, one needs  a well defined 
decoupling limit. For that, since
\be
H_3=\frac{g^2_{D3}N l_S^4}{\bar{r}^4}=\frac{g^2_{D3}N
  l_s^4}{\bar{R}r^3}=\frac{g^2_{D3}N}{U^3}\frac{1}{l_s^4}\frac{l_s^2}{
\bar{R}} 
\label{average}
\ee
one needa to have $R=\frac{l_s^2}{\bar{R}}$ (the T dual radius in the 9d D3
brane) fixed. The D3 brane scalar is identified with 
\be
\phi^9 \sim \phi ^9 +\frac{\bar{R}}{l_s^2}=\phi^9 + \frac{1}{R}
\ee
Then, if $1\gg U l_s$ and $l_s/R \gg U l_s$, one can't see the
identification of $\phi^9$ (we don't probe it). So for R fixed and small 
the field theory is with all the scalars noncompact, and the dual 
of this ${\cal N}$=4 SYM is D3 in 10d. In the regime where one begins 
to probe the D4, we have to have a transverse scalar compact, and 
average over it in the holographic dual as was done in (\ref{average}). 
The D3 field theory should have only one 
holographic dual in a given energy regime, so it is not surprising 
that the compactified D4 holographic dual is not valid if R is
sufficiently small. 
Hence compactifying on a small circle the holographic 
dual forces you to go to the T dual description, even if $R\gg l_s$
(the point being that for transverse coordinates, the distances 
are rescaled by $l_s^2$, so even though $\bar{R}/l_s =l_s/R\ll 1$,
still $1/R=\bar{R}/l_s^2\rightarrow \infty$). 

Since we are interested in a 4d field theory,
we should then look for the T dual description of the D4-D8: 
a D3-D7 system 
smeared over an overall transverse direction. I leave for the 
next section the details of this construction, but from now on I 
will be talking about D3-D7 systems. 

But one still needs to get rid of the scalars (both adjoint and bifundamental)
and the conjugate fermions. The adjoint scalars correspond to the positions 
of the D3 branes, and the bifundamental to the relative D3-D7 positions.

Getting rid of the scalars 
can be done by introducing an effective potential for them,
or equivalently by fixing the motion of the D3 inside the 
D7. It is so since by putting a potential 
for the D3 (by modifying the metric), we change the D3 theory
from pure SYM to a modified SYM dictated by the DBI action in 
that background. For instance, if the D3 is stuck at a metric
singularity, the DBI action will imply a term 
\be
\partial_aX^{\mu}\partial_aX^{\nu}g_{\mu\nu}(X)\rightarrow 0, 
{\rm  as} X\rightarrow X_0
\ee
and so the kinetic term will be null, or by rescaling to a canonical 
form, the potential will be infinitely steep, and thus the
corresponding scalar(s) will dissappear from the spectrum. 

By condensing the $D7-\bar{D7}$ to the D5, one has effectively
insured that the motion of the D3 is fixed in those directions 
(since the D5 position will have a singular metric in its transverse
directions). That means that one still needs to fix the position in 
the directions inside the D5, transverse to the D3. 

A comment is in order here. If there is supersymmetry -e.g. between 
a paralel D3 and a D7 in flat space- then the above argument is not 
true. Anyway, the argument above is for the supergravity
approximation, and we should consider the full string theory. If there is 
susy, there is no potential between D3 and D7. If there is no 
susy (susy broken at the string scale), there will be string scale 
masses for the scalars separating the two. Of course, the D3-D5 
background is still supersymmetric inside the 7-plane, but there 
will be a potential in between the two, since the (3,3) (and (5,5))
fermions are massive, so there will be no cancellation of forces.

But one still needs to fix the position of the D3 inside the D5,
(as well as to get rid of the fundamental fermions in the conjugate 
representation),
so one needs to have a special point inside the D5. It can be either a 
singularity, or a brane.

The simplest way to do that is to put an additional D7', perpendicular 
on the D7. This is a supersymmetric configuration, 
where $D7\perp D7' (5)$, 
such that the common worldvolume with the D5 is the D3. This
configuration still preserves ${\cal N}$=1 supersymmetry, as does the D3-D5-D7
(the D7' doesn't break any additional supersymmetry). Of course, the 
$\bar{D7}$ breaks the susy completely.

When the $D7-\bar{D7}$ condenses to the D5, the ${\cal N}$=1 (D3,D7) 
bifundamental superfield describing the condensation directions will 
dissapear (become massive), so we will be left only with the (D3,D5) 
${\cal N}$ =1 bifundamental superfield. D7' is still needed to make the 
bifundamental scalar massive by giving it a potential as argued above. 

One could correctly argue that by introducing D7' we generate (D3,D7')
fields, but these become massive since their operators couple to fields 
which  can propagate in the whole spacetime, which is not supersymmetric. 
It would seem like we would need to have $N_f \gg N_f'$, in order to treat 
D7' as a perturbation, but the system before the introduction of the 
$\bar{D7}$ is supersymmetric, 
and the $\bar{D7}$ breaks the susy only outside the 
brane, on it is still valid. And in any case, all fermions propagating 
outside the $\bar{D7}$ will become massive. The D3-D7' bifundamental 
scalars are massive for the same reasons that the D3-D7 scalars are. 

One sees now an added reason for going to the D4-D8 system.
 A D8 can't intersect with a 
transverse D8 and still preserve susy, since D8 preserves say
$\Gamma_9\epsilon_0=\pm \epsilon_0$, and a transverse D8 will
preserve say, $\Gamma_8\epsilon_0=\pm \epsilon_0$, but that would 
mean that one needs $[\Gamma_8, \Gamma_9]\epsilon_0=0$ which implies 
$\epsilon_0=0$. A Dp brane can supersymmetrically self-intersect 
over a p-2 brane (as e.g., $D7\perp D7'(5)$), because, e.g.,
$[\Gamma_{89}, \Gamma_{67}]=0$.

Finally, let us address the question of the condensation of the 
$D7-\bar{D7}$ system to D5. There are two ways this can happen. One is 
due to Sen \cite{sentach} and formalized in the language of K theory by 
Witten \cite{wittenk} (for more details see the lectures \cite{sen1}). 
The point is that on a $Dp-\bar{Dp}$ worldvolume 
there is a tachyon field which condenses. At the minimum of its potential, 
the tachyon potential and the tension of the $Dp-\bar{Dp}$ cancel each 
other, $g^{-1}V(T_0)+2T_D=0$ and one has vacuum. But the complex tachyon 
might have a vortex solution which behaves like 
\be
T\simeq T_0 e^{i\theta}, A_{\theta}^{(1)}-A_{\theta}^{(2)}\simeq 1 
\;\; {\rm at}\;\; r\rightarrow \infty
\ee
where the tachyon kinetic term is 
\be
|D_{\mu}T|^2=|\partial_{\mu}-iA_{\mu}^{(1)}+iA_{\mu}^{(2)}T|^2
\ee
and so one gets a magnetic flux for $A_{\mu}^{(1)}-A_{\mu}^{(2)}$ at the core, 
which means that one has a Dp-2 brane. 

The other way by which one could get an Dp-2 endpoint for the condensation 
is if one has an explicit flux from the begining. Either way, one can view the 
supergravity solution as the one before tachyon condensation, when the 
tachyon field is still at T=0 throughout most of the Dp brane, except around 
a Dp-2 core.

Note that, as I mentioned, $g_s$ is small enough to ensure that we can trust 
the supergravity approximation (unlike for other Dp branes, where one needs 
an extra condition, like large N for susy branes and large r for nonsusy 
ones), so even though the system may be unstable, there is an interval of time 
when the $Dp-\bar{Dp}$ solution is correct. And that time should go to 
infinity as $g_s\rightarrow 0$. There was over the last year a flurry of 
activity in the analysis of time dependence on the unstable Dp branes and 
brane-antibrane pair, started by \cite{sen2}. For instance, the analysis 
in \cite{sen3}, done in the context of classical field (and string field)
theory notes that the timescale of quantum effects would go to infinity as 
$g_s\rightarrow \infty$. But in our case the string field theory effects 
themselves ($\alpha '$ effects) will be proportional to $g_s$.

Also note that this mechanism only works for D8 branes (and in general 
smeared Dp branes with only one nontrivial transverse coordinate), since a 
$Dp-\bar{Dp}$ will be a black brane, with a singularity; so for that there 
will be significant stringy corrections. Another way of seeing this is that 
radiating away energy in one transverse dimension is much harder than in 
higher dimensions, so the decay timescale can be actually made infinite.
A further discussion of general $Dp-\bar{Dp}$ brane systems will be done 
in the next subsection. 

So the holographic dual for QCD would have been the $D7\perp D7' (5)$,
with an extra $\bar{D7}$ paralel to the D7, and smearing over one of 
its transverse coordinates, with a D5 remnant intersecting D7' over a 
D3. The problem is that now, we can't write down this solution, not 
even implicitly, but one has at least defined the system.

\subsection{Understanding susy breaking: black holes, Randall-Sundrum,
  compactification and going down to D7 branes}

Why were we able to  write down the $D8-\bar{D8}$ solution? 
In general, if we approach a Dp-brane and a anti-Dp-brane in flat spacetime
(otherwise, there are many nontrivial stable solutions) 
we expect that the $Dp-\bar{Dp}$ brane solution (which will have gravitational
mass, but no charge) will be a uncharged black-p-brane (a
generalization of the Schwarzschild black hole by adding p flat
coordinates). That solution would be
\bea
ds^2&=&-dt^2f(r)+d\vec{x}_p^2 +f^{-1}(r)dr^2 +r^2 d\Omega_n^2\nonumber\\
f(r)&=&  1-\frac{\mu}{r^{n-1}}
\label{bh}
\eea
A charged black p-brane solution will be
\bea
ds^2_s &=& H^{-1/2}_p(-dt^2 f(r)+d\vec{x}_p^2)+H_p^{1/2}(f^{-1}(r)
dr^2+r^2 d\Omega^2)\nonumber\\
e^{\phi}&=&g_s H_p^{\frac{3-p}{4}}, \;\; F_{p+2}=Q_{p+1} vol(\Omega
_{8-p})\nonumber\\
H_p&=& 1+\frac{g_s N l_s^{7-p}}{r^{7-p}}
\eea
But one can also have nontrivial dilaton, and hence have a dilatonic 
``black hole''-type solution. But this solution can only appear in the
presence of a cosmological constant in string frame. That cosmological 
constant is supplied by the constant mass in massive IIA, but can be 
obtained by any constant field strength for a RR field, since the
action is 
\be
S_{string, 10d}=\frac{1}{4 k_{10}^2}
\int d^{10}x\sqrt{g} ( 2e^{-2\phi}R + \tilde{M}^2 
+\sum_p F_{p+2}^2+...)
\ee
The ``dilatonic black hole''
 is the generalization for nontrivial dilaton of the
Randall-Sundrum domain wall inside AdS space 
(positive tension domain wall with nontrivial cosmological
constant). 
The Randall-Sundrum set-up \cite{rs1,rs2} in d dimensions (as opposed to 5)
 has the equations of motion
\be
R_{\mu\nu}-\frac{1}{2}g_{\mu\nu}R=-\lambda g_{ij}\delta(z)
\delta^i_{\mu}\delta^j_{\nu}-\lambda ' g_{\mu\nu}
\ee
which has a solution of the type 
\be
ds^2=A(z)(d\vec{x}^2+dz^2)
\ee
with scalar curvature 
\be
R=\frac{d-1}{A}[(ln A)''+\frac{d-2}{4}((ln A)')^2]
\ee
if 
\be
A=(1-\frac{\lambda}{2(d-2)})^{-2}
\ee
and 
\be
\lambda '=\lambda^2 \frac{d-1}{8(d-2)}
\ee

The dilatonic version of this is obtained in the presence of a constant
field strength and is given by
\bea
ds_s^2&=&H(z)^{-1/2}d\vec{x}^2_{p+1} +H^{1/2}(z)(dz^2+ d\vec{y}_{n})
\nonumber\\
e^{\phi}&=&g_sH^{\frac{3-p}{4}}\nonumber\\
F_{(n)}&=&m\nonumber\\
H&=&1+m|z|
\label{dilbh}
\eea
and this solution breaks supersymmetry and is uncharged, so it is the 
analog of a black hole (or rather, Randall-Sundrum). 

It is interesting to note that RS would break susy anyway (while
keeping susy on the brane), but in string theory one doesn't have a
cosmological constant, but at most constant field strength, as we
saw, which provides a potential (which at constant dilaton can
be interpreted as a cosmological constant). 
 So this solution is as close as one can get to a
Randall-Sundrum type scenario. Also note that (\ref{dilbh}) generates 
$Dp-\bar{Dp}$ solutions for p=4,...,8 (we need a magnetic-type
solution, since $F_{(8-p)}$=ct.), and could be also extended 
for p=3 (with self-dual field strength $F_{(5)}=* F_{(5)}$), and 
this list exhausts all interesting configurations (i.e. configurations
where our 4 dimensions live on the brane). 

Schwarzschild
black holes (non-dilatonic, no cosmological constant) can exist 
in p+1 dimensions, $p>2$, with solution (\ref{bh}). For p=2, we can
still have a solution, but of a different form than (\ref{bh}). For 
convenience, we embed it in 4d as the ``cosmic string'' solution. It
is 
\be
ds^2=-dt^2 +dr^2 +r^2(1-8G\mu)d\theta^2 +dz^2
\ee
The fact that the solution looks different can be understood as the 
manifestation of the fact that 2+1 gravity is of Chern-Simons type, 
hence topological, and thus the black hole is just a conical defect.

However, there are no 1+1 dimensional black holes, the reason being
that the Einstein action is purely topological, $\int R$ is just the 
Euler invariant in 2d, so one can't put a source into the Einstein
equations. There is however a solution for gravity coupled to a 
dilaton, which is just the dimensional reduction of the dilatonic 
black hole. 

This fact implies in 4d the statement that a gravitating infinite 
plane (domain wall) can't have static metric (unlike a cosmic string),
but rather the plane is inflating. The solution (Villenkin) is 
\be
ds^2=-(1-k|z|)^2dt^2+ dz^2 + (1-k|z|)^2 e^{2kt}(dx^2 +dy^2)
\ee

When a higher dimensional black hole solution is compactified on 
a transverse direction, and the radius 
R of the compact dimension y is smaller than the gravitational radius 
$R_G$ of the black hole, then  
\be
f=1-\frac{R_G^m}{(\vec{z}^2 + y^2)^{m/2}}
\ee
gets averaged over y and we get the lower dimensional black hole,
with
\be
\tilde{f}=1-\frac{R_G^m}{z^{m-1}R}
\ee

Let us see in more detail how the $Dp-\bar{Dp}$ solution 
 works for D7, which we saw is our preferred choice.

One would write a D7 solution as 
\bea
ds^2_s&=&H^{1/2}(r, \theta )d\tilde{x}^2
+H^{1/2}(dr^2 +r^2 d\theta^2)\nonumber\\
e^{\phi}&=&H^{-1}, \;\;\; \partial_r a\sim \partial_r H
\eea
and the usual D7 is the oxidation of the stringy cosmic 
string solution, i.e.
\bea
ds_E^2&=& d\tilde{x}^2 +H(r, \theta)(dr^2 +r^2 d\theta^2), \;\;
H=\Omega^2\nonumber\\
H&=&\Omega^2=\tau_2=e^{-\phi}, \;\;\; \tau=\tau (z)\nonumber\\
j(\tau (z))&=& \frac{P(z)}{Q(z)}
\eea
but on the other hand the T dual solution to the D8, which is just the D7 
averaged over one transverse direction, is 
\bea
ds^2_s&=& H^{-1/2}(z)d\tilde{x}^2 +H^{1/2}(z)(dz^2 +dx^2)\nonumber\\
e^{-\phi}&=& H(z), \;\; H(z)=1+m|z|\nonumber\\
a&=&\pm H' x\Rightarrow F_x=\pm H'={\rm ct.}\nonumber\\
\epsilon &=& H^{-1/8}\epsilon_0\;\;\; \Gamma_{zx}\epsilon_0
=\pm i \epsilon_0
\eea
where as usual the $\pm $ refers to D7 versus $\bar{D7}$. We can
easily see that a $D7-\bar{D7}$ solution exists and the only
modification is that it has $F_x=m=$ const., and as a consequence no 
global $\epsilon$. 

One can generalize this D7 solution to the (multiply 
T dualized D8) 
\bea
ds^2_s&=& H^{-1/2}(z)d\tilde{x}_{p+1}^2 +H^{1/2}(z)(dz^2+d\vec{x}_n^2)
\nonumber\\
e^{\phi}&=&H^{\frac{3-p}{4}}(z)\nonumber\\
F_{(n)}&=& \pm H', \;\;\; H(z)=1+m|z|
\eea
and the $Dp-\bar{Dp}$ solution is again obtained by having 
$F_{(n)}=m=$ constant, exactly the ``dilatonic black hole'' 
solution (\ref{dilbh}).

\section{Phenomenology-trying to embed the Standard Mo- del in string 
theory via Dp-Dp+4 systems}

Finally, in this section I will try to see whether we can use the
susy breaking mechanism used for the QCD holographic dual to lift it 
to an embedding of the Standard Model in string theory in the
braneworld approach. I will try to fit the model into a GUT type 
scenario, including SU(5). In the appendix I review GUTs from our 
point of view. The ingredients used for the model building 
are D3-D7-O(7) systems, so I will analyze these first, and move 
to model building in the next subsections.

A few commments are in order about the procedure. The goal is not to 
apply the gravity-gauge duality, but just to take the string theory 
system and lift it to the Standard Model, and see what is obtained 
in the 4d field theory. Since $N_c$ is now small and $\alpha '$ is 
finite but nonzero, on the D3 branes we will have string corrections 
to the YM action. The fact that $N_c$ is small will only affect the 
geometry of the compact space, but I will not make any precise statements 
about that (there will be a significant backreaction). 

As an example of the procedure, 
let's take the ${\cal N}=4$ $SU(N_c)$ SYM - $AdS_5 
\times S_5$ holography and try to make a braneworld model  
(by which in this example I just mean adding gravity and making $N_c$ 
finite). One would make the space transverse to the D3 branes compact.
Since $N_c$ is small, that space would not be approximated by any version 
of $AdS_5 \times S_5$ with the radial AdS direction compactified, but that 
is not what one is after. The corrections to the D3 brane theory (in the 
form of the DBI action) will still 
be small, so the analysis of the field theory should carry through. The 
$1/N_c$ effects will affect the quantitative physics (correlators, etc.), 
but not the qualitative physics (low energy fields and possible
interaction terms). 
The question  deserves further study, but I am going to assume that the 
qualitative physics is unmodified by the small $N_c$.

\subsection{Supersymmetric lagrangeians for D3-D7-O(7) systems}

The Dp-D(p+4) lagrangeian is, dimensionally reduced 
to 4d (the notation used is for D5-D9, but the rest -D3-D7 and 
D4-D8- are the same modulo a relabeling of fields). 
\bea
{\cal L}^{N=2}_{4d}&=&\frac{1}{8\pi}Im Tr [\tau (\int d^2\theta 
W_{\alpha}W^{\alpha}+2\int d^4\theta \Phi^+_i e^{-2V}\Phi_i e^{2V}
\nonumber\\&& +\frac{1}{3!}\int d^2\theta \epsilon_{ijk}\Phi^i\Phi^j\Phi^k
)]_{(5,5)+(9,9)}\nonumber\\&&
+\int d^4\theta(Q^+e^{-2V_{5,5}}Qe^{2V_{9,9}}+\tilde{Q}e^{2V_{5, 5}}
\tilde{Q}^+e^{-2V_{9,9}})\nonumber\\&&
+\int d^2\theta \sqrt{2}(\tilde{Q}\Phi^1_{5,5}Q+\tilde{Q}\Phi^1_{9,9}
Q +h.c.)]
\eea
where by $(5,5)+(9,9)$ we understand the sum over both kind of indices.
The component fields are, as usual:
\bea
V&=&-\theta \gamma^{\mu}\bar\theta A_{\mu}+i\theta^2\bar\theta 
\bar\lambda -i\bar{\theta}^2\theta\lambda +1/2\theta^2\bar{\theta}^2D
\nonumber\\
\Phi^i(y,\theta)&=&A^i(y)+\sqrt{2}\theta\Psi^i(y)+\theta\theta F^i(y)
\nonumber\\
Q&=&a(y)+\sqrt{2}\theta q(y) +\theta\theta f(y)\nonumber\\
\tilde{Q}&=&\tilde{a}(y)+\sqrt{2}\theta\tilde{q}(y)+\theta\theta\tilde{f}
(y)
\eea
and $V$, $\Phi^1=\Phi$ make up an ${\cal N}=2$ vector, whereas $\Phi^2$ and 
$\Phi^3$ make up a hypermultiplet, together making up an ${\cal N}$
=4 vector
 (one for $(5,5)$ fields and one for $(9,9)$ fields), and $Q$ and 
$\tilde{Q}$ make up a hypermultiplet $(5,9)+(9,5)$.

If the D(p+4) theory is decoupled, one just drops the terms involving 
(9,9) fields in  the above. For concreteness, I will talk in D3-D7 
language from now on.

So the superpotential for the ${\cal N}$=2 D3-D7 system  with U(5) 
gauge symmetry is 
\be
{\cal W}=Tr(\epsilon_{ijk}\Phi^i\Phi^j\Phi^k + q^i\Phi^1 \tilde{q}^i)
\ee
and the first term is zero for a U(1) field. The U(1) in U(5)
has a VEV which gives a mass to all the (3,7) hypermultiplet:
$<\Phi^1_{ab}>=m\delta_{ab}$ implies a mass term $mq^i \tilde{q}^i$
for the hypermultiplet, corresponding to separating the D3 and the 
D7 (so $\Phi^1$ correspond to the D3-D7
separation in the overall transverse coordinates). It does not 
give a mass term for $\Phi^2$ and $\Phi^3$, since as we said the 
U(1) piece does not have a superpotential. If we separate 
one D3 from the D7, that would mean giving a VEV to $\Phi^1$, equal 
to $m\delta _{a1}\delta_{b1}$, which does give a mass to the 
$\Phi^{2,3}_{1 a}$ hypermultiplet (strings between D3 brane 1 and the 
rest) and to the $q^i_1$ hypermultiplet (strings between D3 brane 1
and all D7's). Separating one D7 from the rest corresponds to an 
explicit mass term for $q^i$ from the D3 brane theory perspective, 
and comes from the nonzero VEV for $\Phi^{1 (7,7)}_{1 i}$. 

When one goes  to the D3-D7-O(7) system (which is still ${\cal N}$=2
supersymmetric), the 
gauge group becomes Sp(10) (for $N_c=5$ D3 branes), and the superpotential 
becomes
\be
{\cal W}=Tr( W Z'Z +q^i W \tilde{q}^i)
\ee
where ($W_{\alpha}, W$) form a vector in the symmetric of Sp(10) 
(adjoint), (Z', Z) form an antisymmetric hypermultiplet of Sp(10)
and as before $q^i, \tilde{q}^i$ form a fundamental hypermultiplet. W 
extends $\Phi^1$ and Z', Z extend $\Phi^{2,3}$. 

Then the D3-D7-D7' system has ${\cal N}$=1 susy and superpotential
\be
{\cal W}= Tr(\epsilon_{ijk}\Phi^i\Phi^j\Phi^k + q_1^i\Phi^1 \tilde{q}_1^i
+q_2^i \Phi^2 \tilde{q}_2^i)
\ee
where all are ${\cal N}$=1 scalars. The ${\cal N}$=2 
structure was broken by the 
splitting  of the $\Phi^2, \Phi^3$ hypermultiplet into two. 

For the D3-D7-O(7)-D7', the natural guess for the superpotential is
\be
{\cal W}= Tr( W Z'Z +q^i_1W\tilde{q}^i_2 + q_2^i Z' \tilde{q}^i_2)
\ee
(which is the obvious generalization of the D3-D7-D7' case, since the 
presence of the D7' should only be felt in the bifundamental $(q_2^i,
\tilde{q}_2^i)$, coupling to the cooresponding coordinates- Z')
again with obvious meaning. 

Yet another question is what happens when one goes to the 
D3-D7-O(7)-D7'-O(7)' system. The gauge group is (for 5 D3 branes)
$Sp(10) \times Sp(10)$ (a discussion of this model, which is T dual to the 
original Gimon-Polchinski \cite{gipo} orientifold and can be described 
as a $T^4/(Z_2 \times Z_2)$ orientifold, can be found in \cite{sen4,asyt} and 
\cite{afm}). One would guess that the superpotential is 
\be
{\cal W}= Tr( W_a Z'_aZ_a +q^i_1(W_1+Z_2)\tilde{q}^i_2 + 
q_2^i (W_2+Z'_1) \tilde{q}^i_2)
\ee
However, the analysis of \cite{asyt} finds that the fields corresponding to 
coordinates transverse to the D3 brane do not become (W, Z, Z') with 
indices in $Sp(10)\times Sp(10)$ ( $ (W_a, Z_a, Z'_a)$ in the above notation),
but instead become  two chiral multiplets A and B in the (10,10) representation
of the gauge group for the 4 coordinates transverse to D7's ($\Phi^{2,3}$ 
and Z, Z' before), and fields in the (1,1)+(44,1)+(1,1)+(1,44) representation 
for the other two coordinates ($\Phi^1$ and W before). In a $Sp(2k)\times 
Sp(2k)$ theory, it would be (2k,2k) for the first 4 coordinates and 
(1,1) +(k(2k-1)-1,1)+(1,1)+(1,k(2k-1)-1) for the other two (two singlets and 
two antisymmetric traceless representations, one in each Sp factor; for a 
single D3, it would just be two singlets, $S_1$ and $S_2$). 

A simple argument for the gauge group 
is as follows (a more detailed study can be found, e.g. 
in \cite{asyt}). An orientifold makes the open string vector 
wavefunction symmetric in the covering space. With 2 intersecting
O(7) planes, we have 4N D3s in the covering space. Taking into 
account the first O(7) makes the Chan Patton $4N \times 4N$ matrix 
symmetric (symmetric under reflection by the diagonal). The second
O(7) corresponds to rearranging the order of the matrix elements 
and then making it symmetric, or symmetrizing the matrix 
under the second diagonal. By this operation, we are left with 
2N(2N+1) independent elements, enough to form the adjoint of $Sp(2N)
\times Sp(2N)$. The analysis in \cite{asyt} did not find a simple
 superpotential valid everywhere, but for nonzero A and B (and zero 
$S_3$ and $S_4$), it is 
\be
{\cal W}= S_2 AB + q_1^iABq_1^i/\sqrt{A^2} +q_2^iABq_2^i / \sqrt{B^2}
\ee
Here $i=1,..., 2N_f$ (8 at the superconformal point) and $S_2$ is one 
of the two singlets. Presumably one should also have a term involving
$S_3$ and $S_4$, maybe 
\be
{\cal W}= (S_3+S_4)AB
\ee
where $S_3$ and $S_4$ are the antisymmetrics (1,44) and (44,1).
The $Sp(10)\times Sp(10)$ should get broken to a diagonal 
subgroup when O(7)' is removed, and AB probably becomes W (the rest 
of the components become massive). 

\subsection{D3-D7 system; TeV strings}

We have seen that the model for QCD was obtained from a D3-D7-D7' 
system, so that should be a part of the sought after 
Standard Model construction. 
I have also said that we want an SU(5) GUT (most likely embedded into a 
larger GUT), so for the begining, the simplest way to generate an 
SU(5) is by having 5 D3 branes.

So the system we want should have 5 D3 branes and a number of higher
branes, responsible for the fundamental and antisymmetric tensor
 fields.  The gauge group will then be $U(5)=(SU(5) \times U(1))/Z_5$.
If the system separates into 2+3 branes, we have SSB to $U(3) \times 
U(2)$. Out of the remaining 2 U(1)'s, one is the center of mass one
and one is the U(1) inside SU(5). Puting this 5 D3 branes inside a 
$D7-\bar{D7}$, we generate a hypermultiplet in the fundamental (5), 
and 4 real scalars in the adjoint remain (the adjoint fermions become 
massive, as we argued). The hypermultiplet contains 4 real scalars
(2 complex) and 2 complex fermions in the (complex) 5 representation. 
In the SU(5) GUT  we have (see appendix for details) the gauge fields in 
the 24, 3 generations of fermions in the 5 and the $\bar{10}$  (maybe
also a right-handed neutrino singlet per generation) 
and two Higgses, the one responsible 
for  SU(5) breaking in the 24 (real) and the one responsible for
electroweak SSB in the 5 (or the 45, but that's too much)- complex. 
We see that so far we have enough gauge fields and scalars to contain 
this, but we still need more fermions in the 5 and more importantly
fermions in the $\bar{10}$. 

More 5 fermions will be generated by adding more D5s inside the
$D7-\bar{D7}$ and/or more D7's. But how to generate $\bar{10}$s? The 
only way I see is through orientifolds. That will be discussed in the 
next subsection, but let's see how far can we go without O(7)s. 

At this moment one needs to make the following observation. 
In the $D3-D7-\bar{D7}$ type of construction we need the D7 field
theory to be decoupled. That was true exactly only in the case of the $\alpha
'\rightarrow 0$, now we just want it to have a very small
coupling. But we saw that $g^2_{D3}=g_s$=fixed implies $g^2_{Dp}=g_s
l_s^{p-3}$ goes to zero if $l_s\rightarrow 0$. But if one compactifies
on a radius of the order of $l_s$ we are back to square one, since the 
effective 4d gauge theory will have a coupling $g^2_{eff,4d}=g^2_{Dp}/
(\Pi_i R_i)=g_s \Pi_i (l_s/R_i)$. So in order for the Dp  ($p>3$) gauge 
theory to decouple one needs the volume of the extra dimensions
in string units to be very large. If one has just the D3-D5-D7 system, 
one needs only the D5 compactification to be on a large volume, but if
one adds the D7', since the D5 compactification volume is 
transverse to it, one needs to 
have also some of the other coordinates to be very large.

One can put limits on the size of large dimensions  as
follows. 
Whenever one constructs a gauge theory from intersecting branes, 
one gets extra gauge 
fields at the compactification scale, and there are strong
experimental constraints on that. That is why one needs brane
constructions for large extra dimensions: only gravity is 4dimensional 
up to just a mm scale \cite{add}. There are no new gauge bosons up to the
electroweak scale (100 GeV), and depending on their couplings even 
up to the TeV scale \cite{adh}. And when talking 
about unification, anything other than U(1)'s is hard to introduce 
below the GUT
scale. So in conventional scenarios with intersecting branes, where one 
needs the 4d field theory to have finite coupling, the only large
volume that is allowed is transverse to all the intersecting branes. 
A large volume brings down the string scale, and then the energy scale of
new vector bosons can't be too small (the length scale too much 
bigger that the string length). A possible way out of this is by having 
the D brane wrapping cycles be small, but the overall volume of the 
(nontrivial) compact space inhabited by branes be large, but a convincing 
scenario of this type hasn't appeared yet.

Now, the advantage is that one needs small coupling in 4d, so one 
actually {\em needs} large extra dimensions paralel to the D5 and D7'. There
probably is a constraint on how large these can be, but I will not
analyze it.

It is refreshing to see that one also needs large extra 
dimensions from another perspective. We break susy at the string 
scale, so presumably we need to bring down the string scale to TeV, 
while keeping gravity at the Planck scale, which can only be done 
with large extra dimensions. 

Let us recap a few numbers associated with large extra dimensions 
scenarios. The Planck mass in 4d is given by
 $M_{Pl}^2=M^2 (RM)^n$ (n=p-3 is the number of 
large extra dimensions of radius $R\gg 1/M \gg 1/M_{Pl}$. If $M\sim 
10-100$ TeV,  which is the present lower limit for n=2 (coming from cosmology 
and astrophysics),
then $RM= M_{Pl}/M\sim 10^{14-15}$ ($M_{Pl}\sim 10^{19}GeV$), then 
$R\sim 1-100 eV^{-1}\sim .2- 20 \mu m$. If we increase n, we can make R 
even smaller, or equivalently M smaller. 
For instance, with n=4, $R\sim 10^{15/2} M^{-1}\sim 
10^{-5} eV^{-1}\sim 10^{-6}\mu m $.

One must also remember that to get the Standard Model 
one must give masses to the various fermions 
and scalars. One needs to break the SU(5) GUT by giving a 
VEV to an adjoint (24) Higgs. But 
the 5 of SU(5) must remain massless relative to the GUT 
scale $M_{GUT}\gg M_{2,3}$, which can be realized if the 2 and 3 
branes are separated {\em inside} the D5, so that the fundamental
fields remain massless. Then $r_{2,3}\sim l_s^2 M_{GUT}$ ($
<\Phi_{2,3}>\sim M_{GUT}$), so that $r_{2,3}\sim 10^{-3}eV^{-1}\sim
10^{-5}R$ ($r_{2,3}\leq R=$maximum, so it is OK),  so one doesn't need
warping to increase the energy scales (one of the lessons derived from 
the Randall-Sundrum I model \cite{rs1} is that warping dramatically 
increases energy ratios).

\subsection{Adding O(7) planes- how far can we go?}

As I mentioned in the previous subsection, the only way I see to get
a $\bar{10}$ fermion in the above construction is to introduce 
orientifolds. 

The point is that one has a 
$\bar{10}$ in the decomposition of the antisymmetric traceless
 tensor 44, and there 
is an antisymmetric tensor in the D3-D7-O(7) gauge theory. It comes  
from the (3,3) strings , the 4 scalars corresponding to motion inside
the  D7 become a 44 (antisymmetric traceless) hypermultiplet. Unfortunately,
there is only one such hypermultiplet, not $N_{gen}=3$, as we need.
Moreover, it comes from strings stretching between the D3 and its
O(7) image, so when Sp(10) is broken to SU(5) they should become
massive with mass= symmetry breaking scale. One could see the problem 
in another way: the $\bar{10}$ contains a (3,2) fermion , but the 
(3,2) gauge fields are GUT-scale massive. 
The first problem could be solved though by putting an O(7)' plane as 
well, so now we have $S_3$ and $S_4$ in the 44 of each group, but 
consequently also of the diagonal subgroup 
(each multiplet  containing a 10 and a 
$\bar{10}$), and also A and B in the (10,10) each containing a 10 
in the diagonal subgroup, so we do have $N_{gen}=3$ fermions in the $\bar{10}$
among the fields. But short of finding some projection which
eliminates the extra orientifold bosons, while keeping the fermions, 
it is hard to see how to solve the second problem.

One would still need to find a way to give Standard Model fields 
mass. 
The masses to the 2 and the 3 fundamental fermions can be obtained 
by separating the D3 from the D5, while remaining inside the D7. 
With SU(5) restored, this is a mass for the 5 obtained by giving 
a VEV to an adjoint of U(5): $<\phi^1_{ij}>=m\delta_{ij}\rightarrow
{\cal W} =q^i \phi^1 \tilde{q}^i= m q^i \tilde{q}^i$ (giving a VEV 
to the U(1) of U(5), really). From a Higgs perspective, we would 
have expected to give a VEV to a Higgs in the 5 though. That is so 
since the mass term should come from a Yukawa coupling $\bar{\Psi}
H \Psi$, so the Higgs should be in a representation that appears in 
the tensor product of the fermion and antifermion representations.
In the SU(5) GUT, $\Psi$ is in 5 and $\bar{\Psi}$ in the $\bar{10}$, 
and $5 \times \bar{10}=\bar{5}+\bar{45}$. 
But if the Higgs in the $\bar{5}$ is one of the $q^i$'s, and the 
fermion in the $\bar{10}$ comes from a Z field, the superpotential 
term $q^iZ \tilde{q}^i$ does give the required fermion mass.

So let us recap the model so far. We have 5 D3 branes inside the
$D7-\bar{D7} $ condensed to $N_1=N_{gen}=3$ D5s. 
We have added $N_2$ D7's in the
direction transverse to the D5 and paralel to D3.
Up to now the model has a gauge field in the 24 
of SU(5) (the center of mass of the D3 is stuck at the intersection 
of D5 and D7', so the $U(1)_{cm}$is lifted), 
the fermions in the 24 are massive 
and there are still the scalars in the 24. Then there are $N_{gen}=3$
hypermultiplets in the fundamental, each composed of 2 complex
scalars and 2 complex fermions, one in the 5 and one in the
$\bar{5}$. The complex fermions in the 5 must remain massless, the 
ones in the $\bar{5}$ must be massive. This happens for the same reason 
as in the QCD case (the $\bar{5}$ fermions correspond to the directions 
transverse to the D5). 
The complex scalars in the 
5 must become massive, whereas one of the 3 complex scalars in the $\bar{5}$
must be the electroweak Higgs plus its counterpart, so at least the 
electroweak Higgs must be massless.
The GUT Higgs is one of the 4 scalars in the 24, which
means that we must have a separation in a nontrivial direction, and 
all the adjoint scalars have to be massive:

The SU(5) is broken by separating 2 D3s away from the D7' center inside 
the D5 (so that the fundamental fields remain massless). In the absence 
of warping $r_{2,3}\sim 10^{-3}eV^{-1}$ must be (much) smaller than the
radius R of the corresponding direction. For instance, one can choose 
one large dimension inside the D5 and one inside the D7', in which
case one has $R\sim 20\mu m $, which is OK. All the scalars are massive
since the D3 branes are stuck in all directions, except for the 
$q^i$ scalar corresponding to the 2 D3's in the separation direction,
as well as the corresponding adjoint. So one has a massless complex 
doublet (Higgs) and (complex) triplet in the adjoint of SU(2). 
One has the 5 fermions, as  wanted. 
One can generate some masses for the 2 and 3 fermions 
by separating the D3 from the D5, but it is not of the type 
that we want. 

It remains to add the $\bar{10}$s, get masses for the fermions and 
get rid of the triplet scalar. 
We saw that one could add a O(7) and an O(7)' at the respective D7s,
and increase the gauge group to $Sp(10)\times Sp(10)$, but 
in this way one generates
4  fields in the $\bar{10}$ of SU(5) together with many other: the
gauge fields are now in the adjoint of $Sp(10)\times 
Sp(10)$, which
is 110 dimensional, but also there are 2 antisymmetrics in the $44=24+
10 + \bar{10}$ and two bifundamentals in the (10,10). 
Mass terms could come from terms like $q^i Z
\tilde{q}^i$, as noted before (q in the 5, $\tilde{q}$ in the $\bar{5}$, Z in 
the $\bar{10}$). There are unfortunately many fields left
over. A detailed analysis would involve understanding better the effect of the 
anti D-brane $\bar{D7}$ on the action (susy breaking).

All of this is of course, as I mentioned, in the context of 
TeV strings, where unification is renormalized 
in an unknown way by string
theory, and it is not clear how much can be said in the context of 
perturbative physics anyway.

Since one doesn't have a quantitative understanding of the $D7-\bar{D7}$
susy breaking, one can't do much to describe the new physics (susy 
breaking corrections) anyway. All we could do is treat the
${\cal N}$=1 supersymmetric system of D branes, calculate the
superpotential, and qualitatively describe susy breaking and the 
emergence of the Standard Model-like field content.

\section{Discussions and conclusion}

The first result of this paper was the holographic dual for the D4-D8 system, 
given in (\ref{decoupled}). This is different from the holographic
dual of the conformally invariant D4-D8-O(8) system given in
\cite{bo}. The D4-D8 system is nonconformal and a Domain Wall- QFT
correspondence is available, in terms of a 6d SU(2) gauged sugra, 
coupled with a mass parameter h to a U(1) vector multiplet. 
The same (\ref{decoupled}) describes the decoupled $D4-D8-\bar{D8}$ 
if we change the sign of M on one side. In order 
to find a holographic dual to large N QCD, we have to break susy by 
adding a $\bar{D8}$, which together with the D8 will condense to a D6.
The $D4-D8-\bar{D8}-D6$ system is described in (\ref{nondec}) implicitly
(up to one integro-differential equation for the variable $H_{p,q;4}(x)$).
Also, in order to have a 4d field theory, and get chiral fermions, 
we need to go to a D3-D7-D7' system. 

The holographic dual ($l_s\rightarrow 0, N \rightarrow \infty$) 
is then

\begin{tabular}{|l|cccccccccc|} \hline
coord. & 0&1&2&3&4&5&6&7&8&9 \\ \hline
D3 & x&x&x&x&-&-&-&-&-&sm \\ \hline
D7-$\bar{D7}$ & x&x&x&x&x&x&x&x&-&sm \\ \hline
D5 & x&x&x&x&x&x&-&-&-&sm \\ \hline
D7' & x&x&x&x& -&-& x&x&x&x \\ \hline
\end{tabular}

And the fields left over are the SU(N) adjoint gauge field $A_{\mu}^{(ab)}$
and the $(N, N_f)$ fermion $\psi^{ai}$, the ${\cal N}$=1 partner of 
$\phi^{ai}_{4-5}$.

The adjoint fermions are decoupled because their
operators couple to fields moving 
in the nonsupersymmetric bulk, and so get very large anomalous dimensions, 
while the fundamental fermions are still there, since their operators are 
constrained to lie on the supersymmetric D7 plane. The adjoint and fundamental 
scalars get masses because of the D3 being fixed in the extra  dimensions. 
The condensation doesn't take place because it would take a very large time.

The mechanism for susy breaking could be used to embed the Standard 
Model into string theory via a SU(5) (or higher) braneworld GUT model,
 but we find that we come short of that
goal. First of all, one would need a TeV scale string theory scenario, 
which is problematic per se. Without the use of orientifolds, we 
could not find a $\bar{10}$ fermion in the SU(5) GUT scenario. 
With orientifolds, there are too many fields and the masses of the 
Standard Model fields don't seem to be what we want.
Yet it is remarcable that one has a whole new class of nonsupersymmetric
theories similar to the Standard model with the gauge group arising 
on the worldvolume of D branes.

In the context of D6 intersections, remarcable progress has been made towards 
embedding the Standard Model (see \cite{uranga,bailin} for a review). 
In particular, \cite{akrt} gave an embedding of just the Standard Model 
(non susy). One of the versions in \cite{imr} e.g., contains also a $D3-
D7-D7'$ system (but a different version). It is therefore conceivable 
that by combining the virtues of both one could find a good phenomenological 
example of Standard Model embedding. 

The model discussed in the paper (in one possible parametrization) 
is (here l=large, s=small, sm=smeared)

\begin{tabular}{|l|cccccccccc|} \hline
coord. & 0&1&2&3&4&5&6&7&8&9 \\ \hline
2\;D3s \;(A) & x&x&x&x&-&-&-&-&-&sm \\ \hline
3\;D3s \;(B) & x&x&x&x&-&-&-&-&-&sm \\ \hline
D7-$\bar{D7}$ & x&x&x&x&x&x&x&x&-&sm \\ \hline
$N_1$=3 D5s & x&x&x&x&x&x&-&-&-&sm \\ \hline
$N_2$ D7's & x&x&x&x& -&-& x&x&x&x \\ \hline
size &$\infty$&$\infty$&$\infty$&$\infty$& l&s& l&s&s&s \\ \hline
\end{tabular}

The separations are of order (for a simple model) $r_4(AB)\sim 10^{-3}eV^{-1}
\sim r_4(AC)$ ; $(r_4(BC)\sim 0)$. The ``massless'' fields are $q^{ai}_{4-5}
(A)$=Higgs doublet, $\phi_{4-5}^{(ab)}(A)$ (extra complex Higgs triplet) 
and $\psi^{ai}(A,B)$ (fermion matter). When one adds O(7) and O(7)', the 
model becomes quite complicated.

When trying to apply lessons from the holographic dual theory, we have 
relaxed two conditions: $N_c$ is now finite (and small), and $l_s$ is 
nonzero. As a result, gravity is not decoupled, but still lives at a high 
energy scale, but a lot of the arguments go through. In particular, the 
analysis of which fermions decouple (or become string-scale massive, in 
this case), and which fermions remain massless should stay the same. 

It was essential that one had the $Dp-\bar{Dp}$ condensation to Dp-2 
happen in a ``frozen'' gauge theory sector, 
giving the fundamental fields, (as opposed to having the worldvolume 
D3 brane giving the Standard Model gauge theory  as the endpoint of 
condensation), in order to keep the fundamental quarks. It was also 
essential that there was only one nontrivial transverse coordinate
($D8-\bar{D8}$ or rather $D7-\bar{D7}$ with a coordinate averaged 
over, being very small), so that we can have a metastable state 
keeping
the condensation process at bay and still describe what happens in the 
Standard Model field 
theory. Finally, having a D3-D7 system was important, since it allowed 
introduction of another D7' breaking the susy to ${\cal N}$=1, a 
requirement for a good phenomenology (having complex representations 
as opposed to real for ${\cal N}$=2).

{\bf Acknowledgements} I would like to thank Radu Roiban for collaboration
at the initial stages of this project, over the last few years of incubation 
of these ideas, in particular for the nonextremal D4 inside D8 solution, 
the field theory- DW sugra correspondence, and the (5,9) susy lagrangians
and for a critical reading of the manuscript.
I would also like to thank Juan Maldacena for finding errors in previous 
ideas I had about D4-D8 systems, which allowed me to really start this 
project, and also to thank S. Ramgoolam, A. Hanany, W. Taylor for 
discussions. 

\newpage

{\Large\bf{Appendix A. Review of SU(5) GUT and higher GUTs}}

\renewcommand{\theequation}{A.\arabic{equation}}
\setcounter{equation}{0}

I will review now some relevant facts about unification, see e.g. 
\cite{georgi} or \cite{kt}.

The $SU(3) \times SU(2) \times U(1)$
 Standard Model has 3 generations of quarks
and leptons, 
\bea
quarks  \begin{pmatrix} u&\\d &\end{pmatrix} 
&
\begin{pmatrix} c & \\ s & \end{pmatrix} 
&
\begin{pmatrix} t & \\ b & \end{pmatrix} 
\nonumber \\
leptons \begin{pmatrix} e & \\ \nu_e & \end{pmatrix} 
&
\begin{pmatrix} \mu & \\ \nu_{\mu} & \end{pmatrix} 
&
\begin{pmatrix} \tau &\\ \nu_{\tau} & \end{pmatrix}
\eea
together with a Higgs doublet. 

Let us describe the first generation. The creation operators for 
right-handed particles is 
\be
u^+, d^+, e^+, (\bar{u}^+, \bar{d}^+)=\bar{\psi}_i^+, (\bar{e}^+, 
\bar{\nu}^+)=\bar{l}_i^+
\ee
with quantum numbers under $(SU(3), SU(2))_{U(1)_Y}$
\be
u^+: (3,1)_{2/3}, d^+: (3,1)_{-1/3}, e^+:(1,1)_{-1}, 
\bar{\psi}^+: (\bar{3}, 2)_{-1/6}, \bar{l}^+:(1,2)_{1/2}
\ee
and the creation operators for the left handed fields transform in the 
conjugate representation. 
These get unified in SU(5) as follows:
\be
(3,1)_{-1/3} +(1,2)_{1/2}=5
\ee
and
\be
(3,1)_{2/3}+(1,1)_{-1}+(\bar{3}, 2)_{-1/6} =\bar{10}
\ee
So a full generation of quarks and leptons fills up a fundamental (5) 
and a antisymmetric tensor (bar), the $\bar{10}=(\bar{5} \times
\bar{5})_a$, all of which must be massless from the point of view of 
the unification scale. One usually talks about the left-handed operators,
\bea
&& \bar{d}_L, e_L, (\nu _e)_L = \bar{5} \nonumber\\
&& u_L, d_L, \bar{u}_L, \bar{e}_L=10
\eea
The gauge field will be in the adjoint of SU(5), the 24,
and it will contain also the X, Y bosons, the ``leptoquarks''
\bea
&& \begin{pmatrix} Y^{-1/3} &\\ X^{-4/3} &\end{pmatrix}=(3,2)_{-5/3}
\nonumber\\
&&\begin{pmatrix} \bar{X}^{4/3} &\\ \bar{Y}^{1/3} & \end{pmatrix}
=(\bar{3},2)_{5/3}
\eea
which mediate transitions between quarks and leptons and quarks and 
antiquarks, therefore violate B and L, hence give proton decay. 

There must be an adjoint  Higgs (in the 24) which breaks the SU(5).
Indeed, the U(1) generator, S, in SU(5) commutes with 
$SU(3) \times SU(2) \times U(1)$, 
so an adjoint Higgs with VEV in the S direction 
does the trick. It will give mass of the order of the unification 
scale to the X and Y gauge bosons. 

Additionally, one must have a Standard Model Higgs doublet, which 
will give masses to the Standard Model
fermions. To have masses for the u, d, and 
e, this doublet must be contained in either a 5 or a 45 of SU(5). 
So a complex 5 Higgs will do the job (as will a 45). The problem with 
that is though that  the SM doublet must have the mass of a 
few 100 GeV for electroweak SSB, whereas the triplet Higgs $H^{\pm
  1/3}$ can mediate B and L violation, so it must also have the 
mass of the order of the GUT scale 

Yet additionally, if we want a neutrino mass, we need a ``see-saw
mechanism''. It can be added in by a right-handed neutrino,
which in SU(5) can only be a singlet $N_R$. We can 
have a Dirac mass term $m\bar{\nu}_LN_R$, and a large Majorana 
mass term $M N_R N_R$, with total mass terms
\be
\begin{pmatrix} \nu_L &  \bar{N}_R \\ &\end{pmatrix}
\begin{pmatrix} 0 & m \\ m & M \end{pmatrix} 
\begin{pmatrix} \nu_L & \\ \bar{N}_R & \end{pmatrix}
\ee
which can be diagonalized to mass eigenstates of $m_1=m^2/M$ and 
$m_2\simeq M$, with $\nu_1 \simeq \nu_L, \nu_2\simeq N_R$. 

Another unification (most popular at the moment) is given by
 $SO(10)\rightarrow SU(5) \times U(1)$, under which 
the spinor representation splits as $16 \rightarrow 10_{-1} + \bar{5}_3 +
1_{-5}$, the fundamental as $10\rightarrow 5 + \bar{5}$ and the adjoint 
(antisymmetric tensor) splits as $45\rightarrow 24_0 + 10_4 +
\bar{10}_{-4}+1_0$. The advantage is that all the quarks in one 
generation are in one representation (the 16). Moreover, the extra 
1 can be attributed to the righthanded neutrino. The gauge field 
and the SU(5) Higgs live in an adjoint 45, and the Standard Model 
Higgs in a fundamental 10.

Yet another avenue of research is given by the Sp groups obtained 
from orientifolds. In particular, $Sp(10)\rightarrow SU(5) 
\times U(1) = U(5)$, under which  again 
$10\rightarrow 5 + \bar{5}$, and the adjoint (symmetric tensor)
 $55\rightarrow 24 + 15 + \bar{15} +1$ and the antisymmetric 
traceless tensor
$44\rightarrow 24 + 10 + \bar{10}$. The problem is that all these 
representations are real, but we know that symmetry breaking should
work, in the form of the branes going away from the orientifold
point. Note that Sp(2n) has the same rank and the same adjoint
representation as SO(2n+1) (the Sp(2n) adjoint is symmetric=n(2n+1),
and the SO(2n+1) adjoint is antisymmetric= n(2n+1) also, and the 
rank is n for both). 
In particular, Sp(10) will be related to SO(11).

\end{document}